\documentclass[12pt]{article}

\usepackage{setspace}

\usepackage{amsmath}
\usepackage{graphicx}
\usepackage{natbib}
\usepackage{url} 
\usepackage{latexsym}
\usepackage{amsfonts}
\usepackage{amssymb}
\usepackage{psfrag}
\usepackage{graphicx}
\usepackage{calc, cancel}
\usepackage{graphics, graphicx, caption, subcaption}
\usepackage{latexsym, pgfplots, tikz, url,verbatim,multicol,enumerate,bm,listings,color}
\usepackage{multirow}
\usepackage[justification=centering]{caption}

\usepackage{wrapfig, sidecap,enumitem}
\usepackage{epsfig}
\usepackage{longtable}
\usepackage{hyperref}
\usepackage{graphicx}
\usepackage{natbib}
\usepackage{color}
\usepackage{amsgen,amsmath,amstext,amsbsy,amsopn,amssymb}
\usepackage{geometry}
\usepackage{subcaption}
\usepackage{caption}
\usepackage{wrapfig}
\usepackage{fancyhdr}
\usepackage{lastpage}
\usepackage{longtable}
\usepackage{enumitem}

\def\m{\mathbf m}
\def\u{\boldsymbol u}
\def\U{\boldsymbol U}
\def\mV{\mathbf V}

\def\W{\mathbf W}
\def\P{\mathbf P}
\def\Y{\mathbf Y}

\def\w{\boldsymbol w}
\def\y{\boldsymbol y}
\def\p{\mathbf p}
\newcommand{\real}{\text{\rm I\hspace{-0.6mm}R}}

\title{\bf Logistic Normal Multinomial Factor Analyzers for Clustering Microbiome Data}

\author{Wangshu Tu\footnote{Department of Mathematical Sciences, Binghamton University, State University of New York, 4400 Vestal Parkway East, Binghamton, NY, USA 13902. e: wtu2@binghamton.edu} \and Sanjeena Subedi \footnote{Department of Mathematical Sciences, Binghamton University, State University of New York, 4400 Vestal Parkway East, Binghamton, NY, USA 13902. e: sdang@binghamton.edu} }

\date{\today}

\begin{document}
				\doublespacing
	\maketitle
	\begin{abstract}
		The human microbiome plays an important role in human health and disease status. Next generating sequencing technologies allow for quantifying the composition of the human microbiome. Clustering these microbiome data can provide valuable information by identifying underlying patterns across samples. Recently, \cite{fang2020} proposed a logistic normal multinomial mixture model (LNM-MM) for clustering microbiome data. As microbiome data tends to be high dimensional, here, we develop a family of logistic normal multinomial factor analyzers (LNM-FA) by incorporating a factor analyzer structure in the LNM-MM. This family of models is more suitable for high-dimensional data as the number of parameters in LNM-FA can be greatly reduced by assuming that the number of latent factors is small. Parameter estimation is done using a computationally efficient variant of the alternating expectation conditional maximization algorithm that utilizes variational Gaussian approximations. The proposed method is illustrated using simulated and real datasets.
	\end{abstract}
	
	
	\section{Introduction}
The human microbiota is a complex collection of microbes including but not limited to bacteria, fungi, and viruses that reside in the human body. It is estimated that there are nearly 30 trillion bacterial cells living in or on each human body, which is about one bacterium for every cell in the human body \citep{Sender}. These organisms play an important role in human health and diseases \citep{Hutt}. For example, changes in the gut microbiota have been linked to inflammatory bowel disease \citep{Becker}, obesity \citep{Davis}, type 2 diabetes \citep{Cho}, and cancer \citep{Pfirschke}. Using next generating sequencing technologies, the abundance and composition of these microbes can be quantified.

Cluster analysis has been widely used to gain insights from microbiome data. Cluster analysis is used to group observations into homogeneous subpopulations with similar characteristics. Enterotype, a term first proposed by \cite{arumugam2011}, refers to groups of individuals with similar gut microbial communities. \cite{Wu11linkinglongterm} used a partitioning around medoids (PAM) approach with various distance measures to cluster the gut microbiota samples of 98 healthy volunteers and found that the number of enterotypes varied between two and three. \cite{abdel2020} utilized hierarchical clustering to cluster the sputum microbiome datasets and identified two distinct robust phenotype of severe asthma. On the other hand, $k$-means clustering has also been widely used to cluster microbiome data \citep{8593154,hotterbeekx2016}. Although $k$-means, PAM, and hierarchical clustering are well-established clustering techniques and frequently used in many fields, these approaches fail to take into account the compositional nature of the microbiome data. 

Several model-based clustering frameworks have been proposed for microbiome data \citep{DMM2012,subedi2020_JRSSC,fang2020}. A model-based clustering approach utilizes a finite mixture model, which assumes that the data comes from a finite collection of subpopulations or components where each subpopulation can be represented by a distribution function and the appropriate distribution is chosen depending on the nature of the data. A Dirichlet-multinomial model has been widely used for modeling microbiome data \citep{la2012,chen2013,wadsworth2017,koslovsky2020}. In terms of clustering, \cite{DMM2012} proposed a Dirichlet-multinomial mixture model to cluster microbiome data. \cite{subedi2020_JRSSC} proposed mixtures of Dirichlet-multinomial regression models to cluster microbiome data which can incorporate the effects of covariates. However, due to the limited number of parameters in the Dirichlet distribution, the covariance of the microbiome data cannot be modeled adequately using a Dirichlet-multinomial distribution \citep{Xia}. 
 
An alternate model for microbiome data utilized by \cite{Xia} is an additive logistic normal multinomial (LNM) model. An additive logistic normal multinomial (LNM) model \citep{Aitchison} models the observed counts using a hierarchical structure. The observed counts are modeled using a multinomial distribution conditional on the compositions and a Gaussian prior is imposed on the log-ratio transformed compositions. While this approach brings flexibility in modeling the data, the posterior distributions of the transformed variable does not have a closed form solution. A Markov chain Monte Carlo (MCMC) approach is typically utilized for parameter estimation \citep{Xia,aijo2018}, which comes with heavy computational cost. Recently, \cite{fang2020} proposed a mixture of additive logistic normal multinomial (LNM) model to cluster microbiome data and proposed an alternate approach for parameter estimation that utilized variational Gaussian approximations \citep[VGA;][]{wainwright2008}. VGA provides an alternative parameter estimation framework where complex posterior distributions are approximated using computationally convenient Gaussian densities by minimizing the Kullback-Leibler (KL) divergence between the true and the Gaussian densities. 

In the LNM model, the log-ratio transformed composition variable is assumed to be a multivariate Gaussian distribution and hence, the number of parameters in the covariance matrix of the transformed variable grows quadratically with the dimensionality. \cite{McNicholas2008} proposed a family of parsimonious Gaussian mixture models (PGMM) utilizing a factor analyzer structure. In PGMM, the number of parameters in the covariance matrix is linear with dimensionality and by choosing the number of latent factors to be sufficiently small, the number of parameters in the covariance matrix can be greatly reduced. In this paper, we extend the mixture of logistic normal multinomial models for high dimensional data by incorporating a factor analyzer structure in the latent space. We develop a variational variant of the alternating expectation conditional maximization for parameter estimation. The paper is structured as follows: Section \ref{method} provides details of the logistic normal multinomial model and the finite mixture of logistic normal multinomial factor analyzers along with details on parameter estimation; in Sections \ref{sim} and \ref{real}, these models are applied to simulated and real datasets, respectively and Section \ref{conc} concludes the paper.

	\section{Methodology}\label{method}
	\subsection{Additive Logistic Normal Multinomial Model}\label{lnm}
	
	Consider human microbiome count data on K+1 taxa. Let $\mathbf{W}=(W_1,\cdots, W_{K+1})^T$ denote the random vector of counts of K+1 bacterial taxa, and $\p=(p_1, \cdots, p_{K+1})^T$ be the underlying composition of the microbial taxa such that $\sum_{k=1}^{K+1}p_k=1$. The observed counts $\w$ can
be modeled using a multinomial distribution such that
	\[
	f(\w|\p)\propto \prod_{k=1}^{K+1}(p_k)^{w_k}.
	\]
	However, the actual variability in the microbiome composition data is greater than what is modeled or predicted by the multinomial model \citep{Xia}. To account for this additional variability, one approach is to treat the probability vector $\p$ as a random sample from a Dirichlet distribution such that
for each observation $i$, $$\W_i\mid \p_i \sim \text{Multinomial}~(\p_i) ~\text{and}~ \p_i \sim \text{Dirichlet}~(\alpha_1,\ldots,\alpha_{K+1}).$$
The resulting compound distribution is known as the Dirichlet-multinomial distribution and has been used widely for microbiome data \citep{chen2013,DMM2012,subedi2020_JRSSC}. However, due to the limited number of parameters in a Dirichlet-multinomial distribution, the variance and covariances of the microbiome composition cannot be adequately modeled by a Dirichlet-multinomial distribution \citep{Xia}. An alternate approach is to use a log-ratio transformation on $\p$ and impose a prior on the transformed variable \citep{Xia,aijo2018,silverman2018}. 

In this paper, we will use the additive logistic normal multinomial model by \cite{Xia} that utilizes an additive log-ratio (ALR) transformation to map $\mathbf{p}$ from the restricted simplex $\mathbb{S}^K$ to a $K$-dimensional open real space $\real^K$ such that \begin{equation}
\mathbf{Y}=\phi(\mathbf{p})= \left[\log\left(\frac{p_1}{p_{K+1}}\right),\ldots,\log\left(\frac{p_K}{p_{K+1}}\right) \right]^{\top},\label{eqn:Y}
\end{equation} where $p_{K+1}$ is used as a reference and a multivariate Gaussian distribution is imposed with mean $\boldsymbol{\mu}$ and covariance $\boldsymbol{\Sigma}$ on $\mathbf{Y}$. Here, $\phi: (0,1)^{K}\rightarrow \real^{K}$ is a one-to-one function, and therefore,
\begin{equation}
	\p=\phi^{-1}(\Y)=\left[\frac{\exp(Y_1)}{\sum_{k=1}^{K}\exp(Y_k)+1}, \cdots, \frac{\exp(Y_k)}{\sum_{k=1}^{K}\exp(Y_k)+1}, \frac{1}{\sum_{k=1}^{K}\exp(Y_k)+1}\right]^T.\label{eqn:p}
\end{equation}
	Thus, the conditional probability function of $\W\mid\Y$ becomes
	\[
	f(\w|\y)\propto \prod_{k=1}^{K}\left\{\frac{\exp(y_k)}{\sum_{k=1}^{K}\exp(y_k)+1}\right\}^{w_k}\left\{\frac{1}{\sum_{k=1}^{K}\exp(y_k)+1}\right\}^{w_{k+1}},
	\]
and the marginal probability function of $\W$ becomes 
\begin{align*}
f(\w\mid& \boldsymbol{\mu},\boldsymbol{\Sigma})= \int_{\real^K} f( \w| \y)~f(\boldsymbol{ y}|\boldsymbol{\mu},\boldsymbol{\Sigma}_g)~d\y\\
\propto \int_{\real^K} &\prod_{k=1}^{K+1}\left\{{\phi^{-1}(\y)}_k\right\}^{w_k}|\boldsymbol{\Sigma}|^{-\frac{1}{2}}\exp\left\{-\frac{1}{2}(\y-\boldsymbol{\mu}_g)^\top\boldsymbol{\Sigma}^{-1}(\y-\boldsymbol{\mu})\right\} d\y.
\end{align*}
Note that this marginal probability function of $\W$ involves multiple integrals and cannot be further simplified. Although the LNM model provides flexibility in the modeling structure, parameter estimation thus far has mostly relied on Bayesian MCMC-based approaches that come with a heavy computational burden \citep{Xia}. Recently, \cite{fang2020} proposed mixtures of the logistic normal multinomial models (LNM-MM) for clustering microbiome data where a computationally efficient framework for parameter estimation was developed using variational Gaussian approximations \citep[VGA;][]{wainwright2008}. VGA is an alternate parameter estimation framework that utilizes a computationally convenient Gaussian density to approximate a more complex but ``true" posterior density. The complex posterior distribution is approximated by minimizing the Kullback-Leibler (KL) divergence between the true and the approximating densities. 

Using an approximating density $q(\y)$, the marginal log density of $\W $ can be written as:
\begin{align}
	\log f(\boldsymbol{w})&=\int q(\boldsymbol{y})\log\frac{q(\boldsymbol{y})}{f(\boldsymbol{y}|\boldsymbol{w})}d\boldsymbol{y}+\int q(\boldsymbol{y})\log\frac{f(\boldsymbol{w},\boldsymbol{y})}{q(\boldsymbol{y})}d\boldsymbol{y}\nonumber  \\ 
	&= D_{KL}\left[q(\y)||f(\y|\w)\right] + F(q(\y),\w), \label{eqn:loglik}
\end{align}
	where $ D_{KL}\left[q(\y)||f(\y|\w)\right]$ is the Kullback-Leibler divergence from $f(\y|\w)$ to $q(\y)$ and $F(q(\y),\w)$ is known as the evidence lower bound (ELBO). Then, minimizing the Kullback-Leibler divergence from $f(\y|\w)$ to $q(\y)$ is equivalent to maximizing the ELBO. In a variational Gaussian approximation framework,  $q(\y)$ is taken to be a Gaussian distribution. If we assume $q(\y)$ to be a  Gaussian distribution with mean $\m$ and diagonal covariance matrix $\mV$, the lower bound of $F(q(\y),\w)$ becomes
	\begin{equation}
	\begin{split}
	\tilde{F}(\m,\mV,\boldsymbol\mu,\boldsymbol{\Sigma})&=C+{\w^*}^T\m-\left(\sum_{k=1}^{K+1}w_k\right)\left[\log\left(\sum_{k=1}^{K}\exp\left(m_k+\frac{v_k}{2}\right)+1\right)\right]+\\
	\frac{1}{2}\log|\mV|+&\dfrac{K}{2}-\frac{1}{2}\log|\boldsymbol{\Sigma}|-\frac{1}{2}(\boldsymbol{m}-\boldsymbol{\mu})^T\boldsymbol{\Sigma}^{-1}(\boldsymbol{m}-\boldsymbol{\mu})-\frac{1}{2} \text{tr}(\boldsymbol{\Sigma}^{-1}\boldsymbol{\mV}),
	\end{split}\label{eqn:vb}
	\end{equation}
	where $\w^*$ is a $K$-dimensional vector with the first $K$ elements of $\w$ and $C$ is a constant. Details of the derivation of this lower bound is provided in Appendix~\ref{ELBO}. This lower bound can be easily maximized with respect to the model parameters and the variational parameters using an iterative approach. Thus, use of the VGA eliminates the need for an MCMC-based approach for parameter estimation and drastically reduces the computational overhead making it feasible to extend these models for clustering in a high dimensional setting. Several studies have shown that VGA delivers accurate approximations \citep{Archam,arridge2018,chabarber,subedi2020_stat}.


	\subsection{Mixtures of Logistic Normal Multinomial Factor Analyzers}\label{lnmfa}
	A finite mixture model assumes that data comes from a finite collection of subpopulations and each subpopulation can be represented using a parametric distribution. A $G$-component finite mixture of LNM models can be written as
	\[
	f(\w_i|\boldsymbol{\vartheta})=\sum_{g=1}^{G}\pi_gf(\w_i\mid \boldsymbol{\mu}_g,\boldsymbol{\Sigma}_g),
	\]
	where $f(\w_i\mid \boldsymbol{\mu}_g,\boldsymbol{\Sigma}_g)$ represents the marginal probability mass function of the logistic normal multinomial model of the $g^{th}$ subpopulation, $\pi_g>0$ is the mixing proportion of the $g^{th}$ subpopulation such that $\sum_{g=1}^{G}\pi_g=1$, and $\boldsymbol{\vartheta}$ represents all the model parameters. The likelihood of the mixtures of LNM models can be written as 
\begin{equation}
	L(\boldsymbol{\vartheta})=\prod_{i=1}^n\sum_{g=1}^{G}\pi_gf(\w_i\mid \boldsymbol{\mu}_g,\boldsymbol{\Sigma}_g).\label{eq:lik}
\end{equation}
	We define a group membership indicator variable $\boldsymbol{z}_i=(z_{i1},\ldots,z_{iG})$ such that $z_{ig}=1$ if observation $i$ belongs to group $g$ and 0 otherwise. In the context of clustering, these  group memberships are treated as unobserved or missing data and the likelihood function in \ref{eq:lik} is considered an incomplete-data likelihood function.

	Therefore, the complete-data likelihood with observed data ($\mathbf{w}_1,\ldots,\mathbf{w}_n$) and missing data ($\mathbf{z}_1,\ldots,\mathbf{z}_n$) can be written as
	\[
	L(\boldsymbol{\vartheta})=\prod_{i=1}^{n}\prod_{g=1}^{G}\left\{\pi_gf(\w_i|\boldsymbol{\mu}_g,\boldsymbol{\Sigma}_g)\right\}^{z_{ig}}.
	\]
	Then, the complete-data log-likelihood becomes
		\[
	l(\boldsymbol{\vartheta})=\sum_{i=1}^{n}\sum_{g=1}^{G}z_{ig}\left\{\log \pi_g+ \log f(\w_i|\boldsymbol{\mu}_g,\boldsymbol{\Sigma}_g)\right\}.
	\]
	For incorporating a factor analyzer structure \citep{Ghahramani97theem,McLachlan00mixturesof} in the mixtures of LNM models, we utilize the following structure on $\mathbf{Y}$ from the $g^{th}$ component:
\begin{equation*}
	\boldsymbol{Y}=\boldsymbol{\mu}_g+\boldsymbol{\Lambda}_g\boldsymbol{U}_{g}+\epsilon_{g},
\end{equation*}
	where $\boldsymbol{\mu}_g$ is a $K$-dimensional mean vector, $\boldsymbol{U}_{g}\sim N(0,\mathbf{I}_{q})$ is $q$-dimensional vector of latent factors, $\boldsymbol{\Lambda}_g$ is a $K\times q$ matrix of factor loadings, $\epsilon_{g}\sim N(0, \mathbf{D}_g)$ is a $K$-dimensional vector of errors where $\mathbf{D}_g$ is diagonal matrix, and $\boldsymbol{U}_{g}\perp\epsilon_{g}$. Thus, for the $g^{th}$ component, $\boldsymbol{Y}\sim N(\boldsymbol{\mu}_g, \boldsymbol{\Lambda}_g\boldsymbol{\Lambda}_g^T+\boldsymbol{D}_g)$ and $\boldsymbol{Y}\mid \boldsymbol{u}_{g}\sim N(\boldsymbol{\mu}_g+\boldsymbol{\Lambda}_g\boldsymbol{u}_{g}, \boldsymbol{D}_g)$.

	\subsection{Parameter Estimation}
Parameter estimation of the mixtures of factor analyzers is typically done using an alternating expectation conditional maximization (AECM) algorithm. The AECM algorithm \citep{AECM} is an extension of the expectation-maximization (EM) algorithm \citep{dempster77} that uses different specification of missing data at different cycles and the maximization step comprises of a series of conditional maximizations. Each cycle of the AECM algorithm consists of an E-step in which the expected value of the complete-data log-likelihood is computed, which is then followed by a conditional maximization step where a subset of the model parameters are updated. Here, we will develop a variational version of the AECM algorithm that uses different specification of the missing data at different cycles.

\noindent \underline{\textbf{First Cycle}}\\
In the first cycle, we utilize the following hierarchical structure:
$$ \W_i\mid \Y_i \sim \text{Multi.}(\p_i) \quad \text{and} \quad \Y_i\sim N(\boldsymbol{\mu}_g,\boldsymbol{\Lambda}_g\boldsymbol{\Lambda}_g^T+\boldsymbol{D}_g),$$
where $\p_i$ can be obtained from $\Y_i$ using Equation \ref{eqn:p}. Then the component specific marginal probability function of the observed data $\w_i$ is
\begin{align*}
f(\w_i\mid& \boldsymbol{\mu}_g,\boldsymbol{\Lambda}_g,\boldsymbol{D}_g)= \int_{\real^K} f( \w_i| \y_i)~f(\boldsymbol{ y}_i|\boldsymbol{\mu}_g,\boldsymbol{\Lambda}_g\boldsymbol{\Lambda}_g^T+\boldsymbol{D}_g)~d\y\\
\propto \int_{\real^K} &\prod_{k=1}^{K+1}\left\{{\phi^{-1}(\y_i)}_k\right\}^{w_k}|\boldsymbol{\Lambda}_g\boldsymbol{\Lambda}_g^T+\boldsymbol{D}_g|^{-\frac{1}{2}}\exp\left\{-\frac{1}{2}(\y_i-\boldsymbol{\mu}_g)^\top(\boldsymbol{\Lambda}_g\boldsymbol{\Lambda}_g^T+\boldsymbol{D}_g)^{-1}(\y_i-\boldsymbol{\mu}_g)\right\} d\y.
\end{align*}

 Assuming $\mathbf{Z}$ and $\Y$ as missing variables, the complete-data log-likelihood using the marginal probability function of $\W$ is
	\begin{align*}
	l(\boldsymbol{\vartheta}|\w_i)&=\sum_{i=1}^{n}\sum_{g=1}^{G}z_{ig}\left\{\log \pi_g+ \log f(\w_i|\boldsymbol{\mu}_g,\boldsymbol{\Lambda}_g\boldsymbol{\Lambda}_g^T+\boldsymbol{D}_g)\right\} \\
	&= \sum_{i=1}^{n}\sum_{g=1}^{G}z_{ig}\left\{\log \pi_g +\log\int f(\mathbf w_i|\mathbf y_i)f_g(\boldsymbol{ y}_i|\boldsymbol{\mu}_g,\boldsymbol{\Lambda}_g\boldsymbol{\Lambda}_g^T+\boldsymbol{D}_g)d\y\right\}.
\end{align*}

Assuming the component-specific $q(\y)$ to be a Gaussian distribution with mean $\m_g$ and diagonal covariance matrix $\mV_g$ and replacing the log of the marginal of the component probability function by the component specific $\tilde{F} (\m_{ig},\mV_{ig},\boldsymbol{\mu}_g,\boldsymbol{\Sigma}_g)$, the variational Gaussian lower bound of complete-data log-likelihood can be written as
	\begin{equation*}
	\begin{split}
	\tilde{\mathcal{L}}_1 =& \sum_{i=1}^{n}\sum_{g=1}^{G} z_{ig}\left\{\log \pi_g -\left(\boldsymbol{1}_{(K+1)}^T\w_{i}\right)\left[\log\left(\boldsymbol{1}_{(K)}^T\exp\left(\m_{ig}+\frac{\text{diag}(\mV_{ig})}{2}\right)+1\right)\right]\right.\\
	&+ C_i+{\w^*}_i^T\m_{ig}+\frac{1}{2}\log|\mV_{ig}|+\dfrac{K}{2}-\frac{1}{2}\log|\boldsymbol{\Sigma}_g|-\frac{1}{2} \text{tr}(\boldsymbol{\Sigma}_g^{-1}\boldsymbol{\mV}_{ig})\\
	&\left.-\frac{1}{2}(\boldsymbol{m}_{ig}-\boldsymbol{\mu}_g)^T\boldsymbol{\Sigma}_g^{-1}(\boldsymbol{m}_{ig}-\boldsymbol{\mu}_g)\right\},
	\end{split}
	\end{equation*}
	where $\boldsymbol{1}_{(K)}$ stands for column vector of 1's with dimension $K$, $C_i$ stands for $\log\frac{\boldsymbol{1}^T\w_{i}!}{\prod_{k=1}^{K}\w_{ik}!}$, $ \text{diag}(\mathbf{V}_{ig}) =(\mathbf{V}_{ig,11},\mathbf{V}_{ig,22},\ldots,\mathbf{V}_{ig,KK})$ puts the diagonal elements of the $K \times K$ matrix $\mathbf{V}_{ig}$ into a K-dimensional vector, and $\boldsymbol{\Sigma}_g=\boldsymbol{\Lambda}_g\boldsymbol{\Lambda}_g^T+\boldsymbol{D}_g$. In this cycle, for the parameter updates in the $(t+1)^{th}$ iteration, the following steps are conducted: 
	\begin{enumerate}
		\item Update the variational Gaussian lower bound of the complete-data log-likelihood from the first cycle $\tilde{\mathcal{L}}_1$ by updating $\m_{ig}$ and $\mV_{ig}$. For updating $\mV_{ig}^{(t+1)}$, we use the Newton-Raphson method. We take the derivative respect to standard error $v_{ig}^{(t+1)}$ and find the solution to the following score function:
		\[
		\frac{\partial \tilde{\mathcal{L}}_1}{\partial v_{ig}}={v_{ig}^{(t)}}^{-1}-v_{ig}^{(t)}\text{diag}({\boldsymbol{\Sigma}_g^{(t)}}^{-1})-(\boldsymbol{1}_{(K+1)}^T\w_{i})v_{ig}^{(t)}\frac{\text{diag}(\exp(\m_{ig}^{(t)}+\frac{\text{diag}(v_{ig}^{(t)})^2}{2}))}{\boldsymbol{1}_{(K)}^T\exp(\m_{ig}^{(t)}+\frac{\text{diag}(v_{ig}^{(t)})^2}{2})+1}.
		\]
		
		For updating $\m_{ig}^{(t+1)}$, we again use the Newton-Raphson method to find the solution to the following score function:
	\[
	\frac{\partial \tilde{\mathcal{L}}_1}{\partial\m_{ig}}=\w^*_i-{\boldsymbol{\Sigma}_g^{(t)}}^{-1}(\m_{ig}^{(t)}-\boldsymbol{\mu}_g^{(t)})-(\boldsymbol{1}_{(K+1)}^T\w_{i})\frac{\exp\left(\m_{ig}^{(t)}+\frac{\text{diag}(v_{ig}^{(t)})^2}{2}\right)}{\boldsymbol{1}_{(K)}^T\exp(\m_{ig}^{(t)}+\frac{\text{diag}(v_{ig}^{(t)})^2}{2})+1}.
	\]

	\item Update the component indicator variable $Z_{ig}$. Conditional on the variational parameters $\mathbf{m}_{ig}^{(t+1)}$, $\mathbf{V}_{ig}^{(t+1)}$ and on $\boldsymbol{\mu}_g^{(t)}$, $\boldsymbol{\Lambda}_g^{(t)}$, and $\mathbf{D}_g^{(t)}$, the expected value of $Z_{ig}$ can be computed as
	\[ E(Z_{ig}^{(t+1)})=\frac{\pi_g^{(t)} f(\w_i\mid \boldsymbol{\mu}_g^{(t)}, \boldsymbol{\Lambda}_g^{(t)}, \mathbf{D}_g^{(t)})}{\sum_{h=1}^G \pi_h^{(t)} f(\w_i\mid \boldsymbol{\mu}_h^{(t)}, \boldsymbol{\Lambda}_h^{(t)}, \mathbf{D}_h^{(t)})}.\]
	As this involves the marginal distribution of $\W$, which is difficult to compute, we use an approximation of $E(Z_{ig}^{(t+1)})$ using the ELBO:	
	\[\hat{z}_{ig}^{(t+1)}=\frac{\pi_g^{(t)}\exp\{\tilde{F}(\boldsymbol{\mu}_g^{(t)}, \boldsymbol{\Lambda}_g^{(t)}{\boldsymbol{\Lambda}_g^{(t)}}^T+\boldsymbol{D}_g^{(t)}, \boldsymbol{m}_{ig}^{(t+1)}, \boldsymbol{V}_{ig}^{(t+1)})\}}{\sum_{g=1}^{G}\pi_g^{(t)} \exp\{\tilde{F}(\boldsymbol{\mu}_g^{(t)}, \boldsymbol{\Lambda}_g^{(t)}{\boldsymbol{\Lambda}_g^{(t)}}^T+\boldsymbol{D}_g^{(t)}, \boldsymbol{m}_{ig}^{(t+1)}, \boldsymbol{V}_{ig}^{(t+1)})\}}.\] 

	\item Given the variational parameters and $\hat{z}_{ig}^{(t+1)}$, we then update the parameters $\pi_g$ and $\boldsymbol{\mu}_g$ as:
\begin{align*}
	\hat{\pi}_g^{(t+1)}=\frac{\sum_{i=1}^{n}\hat{z}_{ig}^{(t+1)}}{n}, \quad \text{and} \quad \hat{\boldsymbol{\mu}}_g^{(t+1)}=\frac{\sum_{i=1}^{n}\hat{z}_{ig}^{(t+1)}\boldsymbol{m}_{ig}^{(t+1)}}{\sum_{i=1}^{n}\hat{z}_{ig}^{(t+1)}}.
\end{align*}
	\end{enumerate}
	
\noindent \underline{\textbf{Second Cycle}}\\
In the second cycle, we utilize the following hierarchical structure:
$$ \W_i\mid \Y_i \sim \text{Multi.}(\P_i), \quad \Y_i\mid \U_i=\u_i \sim N(\boldsymbol{\mu}_g,+\boldsymbol{\Lambda}_g\u_{i},\boldsymbol{D}_g),\quad \text{and} \quad \U_i \sim N(\mathbf{0},\mathbf{I}_q),$$
where $\P_i$ can be obtained from $\Y_i$ using Equation \ref{eqn:p}. Assuming $\mathbf{Z}$, $\Y$ and $\U$ as missing variables, the complete log-likelihood using the marginal probability function of $\W$ has the following form: 
	\begin{equation*}
	\begin{split}
	l_2(\boldsymbol{W}, \boldsymbol{Z})=\sum_{i=1}^{n}\sum_{g=1}^{G}z_{ig}\left\{\log\pi_g+\log\left[\int f(\boldsymbol{w}_i|\boldsymbol{y}_i)f_g(\boldsymbol{y}_i|\boldsymbol{\mu}_g+\boldsymbol{\Lambda}_g\boldsymbol{u}_{i}, \boldsymbol{D}_g)f_g(\boldsymbol{u}_{i}|0, \mathbf{I}_{q})~d\y ~d\u \right]\right\}
	\end{split}
	\end{equation*}
	
In this cycle, we derive an approximate lower bound for the log of the marginal probability function of $\W$ using the approximating density $q(\y, \u)$
\begin{align}
	\log f(\boldsymbol{w})&=\int q(\boldsymbol{y},\u)\log\frac{q(\boldsymbol{y},\u)}{f(\boldsymbol{y},\u |\boldsymbol{w})}d\boldsymbol{y}~d\u+\int q(\boldsymbol{y},\u)\log\frac{f(\boldsymbol{w},\u,\boldsymbol{y})}{q(\boldsymbol{y},\u)}d\boldsymbol{y}~d\u\nonumber  \\ 
	&= D_{KL}\left[q(\y,\u)||f(\y,\u|\w)\right] + F(q(\y,\u),\w), \label{eqn:loglik}
\end{align}
where $F(q(\y,\u),\w)$ is the ELBO and $D_{KL}\left[q(\y,\u)||f(\y,\u|\w)\right]$ is the Kullback-Leibler divergence from $f(\y,\u|\w)$ to $q(\y,\u)$. Furthermore, assuming $q(\u,\y)=q(\u)q(\y)$, $q(\u)= N(\tilde{\m}_{ig},\tilde{\mV}_{g})$, and $q(\y)= N(\m_{ig},\mV_{ig})$, we can show that
	\begin{equation*}
	\begin{split}
	F(q(\u,\y),\w)&\geq C+{\w_i^*}^T\m_{ig}-\left(\boldmath{1}^T_{(K+1)}\w_i\right)\left[\log\left(\boldmath{1}^T_{(K)}\exp\left(\m_{ig}+\frac{\mV_{ig}}{2}\right)+1\right)\right]\\
	&+\frac{1}{2}(\log |\mV_{ig}|+\log|\tilde{\mV}_{g}|+q+K-\log|\boldsymbol{D}_g|-\tilde{\m}_{ig}^T\tilde{\m}_{ig}-\text{tr}(\tilde{\mV}_g)\\
	&-\text{tr}(\boldsymbol{D}_g^{-1}(\boldsymbol{V}_{ig}+(\boldsymbol{m}_{ig}-\boldsymbol{\mu}_g)^T(\boldsymbol{m}_{ig}-\boldsymbol{\mu}_g)))+2(\boldsymbol{m}_{ig}-\boldsymbol{\mu}_g)^T\boldsymbol{D}_g^{-1}\boldsymbol{\Lambda}_g\tilde{\m}_{ig}\\
	&-\tilde{\m}_{ig}^T\boldsymbol{\Lambda}_g^T\boldsymbol{D}_g^{-1}\boldsymbol{\Lambda}_g\tilde{\m}_{ig}-tr(\boldsymbol{\Lambda}_g^T\boldsymbol{D}_g^{-1}\boldsymbol{\Lambda}_g\tilde{\mV}_g))\\
	&=\tilde{F}_2(\boldsymbol{\mu}_g, \boldsymbol{\Lambda}_g,\boldsymbol{D}_g,\m_{ig},\mV_{ig},\tilde\m_{ig},\tilde\mV_{g}).
	\end{split}
	\end{equation*}
	Here, $\m_{ig}$ and $\mV_{ig}$ are the variational parameters of $q(\y_{i})$ from first cycle and $\tilde{\m}_{ig}$ and $\tilde{\mV}_{ig}$ are the variational parameters of $q(\u_{i})$. Details of the derivation of the lower bound are provided in Appendix \ref{cycle2}. In this cycle, for the parameter updates in the $(t+1)^{th}$ iteration, the following steps are conducted: 
		\begin{enumerate}
			\item Update the variational Gaussian lower bound of complete-data log-likelihood of the second cycle $\tilde{\mathcal{L}}_2$ by updating  $\tilde{\m}_{ig}^{(t+1)}$ and $\tilde{\mV}_{g}^{(t+1)}$ as
\begin{align*}
	\tilde{\m}_{ig}^{(t+1)}&=({\boldsymbol{\Lambda}_g^{(t)}}^T{\boldsymbol{D}_g^{(t)}}^{-1}\boldsymbol{\Lambda}_g^{(t)}+\mathbf{I}_{q})^{-1}{\boldsymbol{\Lambda}_g^{(t)}}^T{\boldsymbol{D}_g^{(t)}}^{-1}(\boldsymbol{m}_{ig}^{(t+1)}-\boldsymbol{\mu}_g^{(t+1)}), ~\text{and} \\
	\tilde{\mV}_{g}^{(t+1)}&=({\boldsymbol{\Lambda}_g^{(t)}}^T{\boldsymbol{D}_g^{(t)}}^{-1}\boldsymbol{\Lambda}_g^{(t)}+\mathbf{I}_{q})^{-1}.
\end{align*}
		\item Update the group indicator variable $\mathbf{Z}$. Similar to the first cycle, we compute an approximation of $E(Z_{ig})$ using the ELBO from the second cycle:	
	\[\hat{z}_{ig}^{(t+1)}=\frac{\pi_g^{(t+1)}\exp\{\tilde{F}_2(\boldsymbol{\mu}_g^{(t+1)}, \boldsymbol{\Lambda}_g^{(t)},\boldsymbol{D}_g^{(t)}, \boldsymbol{m}_{ig}^{(t+1)}, \boldsymbol{V}_{ig}^{(t+1)},\tilde{\m}_{ig}^{(t+1)},\tilde{\mV}_{g}^{(t+1)})\}}{\sum_{h=1}^{G}\pi_h^{(t+1)} \exp\{\tilde{F}_2(\boldsymbol{\mu}_h^{(t+1)}, \boldsymbol{\Lambda}_h^{(t)},\boldsymbol{D}_h^{(t)}, \boldsymbol{m}_{ih}^{(t+1)}, \boldsymbol{V}_{ih}^{(t+1)},\tilde{\m}_{ih}^{(t+1)},\tilde{\mV}_{h}^{(t+1)})\}}.\] 

\item Update ${\boldsymbol{D}_g^{(t+1)}}^{-1}$ and $\boldsymbol{\Lambda}_g^{(t+1)}$ as
\begin{align*}
	\hat{\boldsymbol{D}}_g^{(t+1)}&=\text{diag}\{\hat{\boldsymbol{\Sigma}}_g^{(t+1)}-2\boldsymbol{\Lambda}_g^{(t)}({\boldsymbol{\Lambda}_g^{(t)}}^T{\boldsymbol{D}_g^{(t)}}^{-1}\boldsymbol{\Lambda}_g^{(t)}+\mathbf{I}_{q})^{-1}{\boldsymbol{\Lambda}_g^{(t)}}^T{\boldsymbol{D}_g^{(t)}}^{-1}\hat{\boldsymbol{S}}_g^{(t+1)}+\boldsymbol{\Lambda}_g^{(t)}\boldsymbol{\theta}_g^{(t+1)}{\boldsymbol{\Lambda}_g^{(t)}}^T\},\\
	\hat{\boldsymbol{\Lambda}}_g^{(t+1)}&=\hat{\boldsymbol{S}}_g^{(t+1)}{\boldsymbol{\beta}_g^{(t+1)}}^T{\boldsymbol{\theta}_g^{(t+1)}}^{-1},
	\end{align*}
where 
\begin{align*}
	\hat{\boldsymbol{S}}_g^{(t+1)}&=\frac{\sum_{i=1}^{n}z_{ig}^{(t+1)}(\boldsymbol{m}_{ig}^{(t+1)}-\boldsymbol{\mu}_g^{(t+1)})^T(\boldsymbol{m}_{ig}^{(t+1)}-\boldsymbol{\mu}_g^{(t+1)})}{\sum_{i=1}^{n}\hat{z}_{ig}^{(t+1)}},\\
	\hat{\boldsymbol{\Sigma}}_g^{(t+1)} &= \dfrac{\sum_{i=1}^n z_{ig}^{(t+1)} \left[\boldsymbol{V}_{ig}^{(t+1)}+(\boldsymbol{m}_{ig}^{(t+1)}-\boldsymbol{\mu}_g^{(t+1)})(\boldsymbol{m}_{ig}^{(t+1)}-\boldsymbol{\mu}_g^{(t+1)})^\top\right]}{\sum_{i=1}^{n}\hat{z}_{ig}^{(t+1)}},\\
\boldsymbol{\theta}_g^{(t+1)}&=({\boldsymbol{\Lambda}_g^{(t)}}^T{\boldsymbol{D}_g^{(t)}}^{-1}\boldsymbol{\Lambda}_g^{(t)}+\mathbf{I}_{q})^{-1}+\boldsymbol{\beta}_g^{(t+1)}\boldsymbol{S}_g^{(t+1)}{\boldsymbol{\beta}_g^{(t+1)}}^T, ~ \text{and} \\
\boldsymbol{\beta}_g^{(t+1)}&=({\boldsymbol{\Lambda}_g^{(t)}}^T{\boldsymbol{D}_g^{(t)}}^{-1}\boldsymbol{\Lambda}_g^{(t)}+\mathbf{I}_{q})^{-1}{\boldsymbol{\Lambda}_g^{(t)}}^T{\boldsymbol{D}_g^{(t)}}^{-1}.
	\end{align*}	
	\end{enumerate}
	
				Overall, our algorithm consists of the following steps:
	\begin{enumerate}[label=\Roman*.]
		\item Specify the number of clusters: $G$ and $q$ and provide an initial guesses for $\boldsymbol{\Lambda}_g, \boldsymbol{D}_g$ and $Z_{ig}$. 
		
		\item  First cycle:
		\begin{enumerate}[label=\arabic*)]
			\item Update the variational Gaussian lower bound of complete-data log-likelihood of the first cycle by estimating $\mV_{ig}$ and $\m_{ig}$. 
			\item  Update $Z_{ig}$.
			\item Update $\pi_g$ and $\boldsymbol{\mu}_g$.
		\end{enumerate}
		\item Second cycle:
		\begin{enumerate}[label=\arabic*)]
			\item Update the variational Gaussian lower bound of complete-data log-likelihood of the first cycle by estimating $\tilde{\mV}_{ig}$ and $\tilde{\m}_{ig}$. 
			\item Update $Z_{ig}$ again.
			\item Update $\boldsymbol{S}_g, \boldsymbol{\Sigma}_g, \boldsymbol{D}_g$, and $\boldsymbol{\Lambda}_g$.
		\end{enumerate}
		\item Compute the likelihood $\sum_{i=1}^{n}\log\sum_{g=1}^{G}\pi_gf(\w_i|\vartheta_g)$ using the current estimators and check for convergence. If it is converged, then stop, otherwise go to step 2. 
	\end{enumerate}
	
		Note that $\boldsymbol{\Lambda}_g$ is unidentifiable. This can be seen if we let $\boldsymbol{\Lambda}_g^*=\boldsymbol{\Lambda}_g\mathbf{T}$ be a new factor loading matrix where $\mathbf{T}$ be an orthonormal matrix such that $\mathbf{T}\mathbf{T}^T =\mathbf{I}$, then $\boldsymbol{\Lambda}_g^*\boldsymbol{\Lambda}_g^{*^T}+\mathbf{D}_g=\boldsymbol{\Lambda}_g\mathbf{T}\mathbf{T}^T\boldsymbol{\Lambda}_g^T+\mathbf{D}_g=\boldsymbol{\Lambda}_g\boldsymbol{\Lambda}_g^T+\mathbf{D}_g=\boldsymbol{\Lambda}_g\boldsymbol{\Lambda}_g^T+\mathbf{D}_g$. Hence, we focus on the recovery of $\boldsymbol{\Sigma}_g=\boldsymbol{\Lambda}_g\boldsymbol{\Lambda}_G^T+\mathbf{D}_g$ which is identifiable. Additionally, by incorporating a factor structure, we can utilize Woodbury Identity\citep{woodbury1950} to compute $\boldsymbol{\Sigma}_g^{-1}$:
	\[\boldsymbol{\Sigma}_g^{-1}=(\boldsymbol{\Lambda}_g\boldsymbol{\Lambda}_g^T+\boldsymbol{D}_g)^{-1}=\mathbf{D}_g^{-1}-\mathbf{D}_g^{-1}\boldsymbol{\Lambda}_g(\mathbf{I}_q+\boldsymbol{\Lambda}_g^T\mathbf{D}_g^{-1}\boldsymbol{\Lambda}_g)^{-1}\boldsymbol{\Lambda}_g^T\mathbf{D}_g^{-1},\]
	and thus the matrix inversion is $O(q^3)$ as opposed to $O(K^3)$. Therefore, when $q\ll K$, inverting $\boldsymbol{\Sigma}$ is computationally efficient.

	\subsection{A Family of Mixture Model for Clustering}
	To introduce parsimony, we further imposed constraints on the parameters of the covariance matrix of the latent variable across groups such that $\boldsymbol{\Lambda}_g=\boldsymbol{\Lambda}$ and $\boldsymbol{D}_g=\boldsymbol{D}$ and on whether $\boldsymbol{D}_g=d_g\mathbf{I}$. This results in a family of eight different parsimonious LNM-FAs (Table \ref{p}). Similar constraints on the components of the covariance matrices were utilized by \cite{McNicholas2008,subedi2013,subedi2015}.  
	
	\begin{table}[!htbp]
		\scriptsize
		\centering
		\caption{The family of logistic normal multinomial factor analyzers.}\label{p}
		\begin{tabular}{@{\extracolsep{5pt}}lcccc}
			\\[-1.8ex]\hline
			\hline \\[-1.8ex]
			\multicolumn{1}{c}{Model} &  \multicolumn{1}{c}{$\boldsymbol{\Lambda}_g$}&\multicolumn{2}{c}{$\boldsymbol{D}_g$}&\multicolumn{1}{c}{Total Par}\\
			\cline{3-4}		\\[-1.8ex]
			&Group&Group& Diagonal&\\
			\hline\\[-1.8ex]
			``UUU" & U&U&U&$G*(Kq-q(q-1)/2)+K*G+G-1+K$\\
			``UUC"&U&U&C&$G*(Kq-q(q-1)/2)+G+G-1+K$\\
			``UCU"&U&C&U&$G*(Kq-q(q-1)/2)+K+G-1+K$\\
			``UCC"&U&C&C&$G*(Kq-q(q-1)/2)+1+G-1+K$\\
			``CUU" & C&U&U&$Kq-q(q-1)/2+K*G+G-1+K$\\
			``CUC"&C&U&C&$Kq-q(q-1)/2+G+G-1+K$\\
			``CCU"&C&C&U&$Kq-q(q-1)/2+K+G-1+K$\\
			``CCC"&C&C&C&$Kq-q(q-1)/2+1+G-1+K$\\
			\hline \\[-1.8ex]
		\end{tabular}
	\end{table}
	In Table \ref{p}, the column ``Group" refers to constraints across groups, the column ``Diagonal" refers to the matrix having the same diagonal elements, and the letters refer to whether or not the constraints were imposed on the parameters such that ``U" stands for unconstrained and ``C" stands for constrained. 
For example, the model ``UCU" refers to unconstrained $\boldsymbol{\Lambda}_g$ but constrained $\boldsymbol{D}_g=\boldsymbol{D}$. Whereas in the model ``UCC" where constraints on both the ``Group" and the ``Diagonal" are imposed for $\boldsymbol{D}_g$, it means $\boldsymbol{D}_g=d\boldsymbol{I}_p$. Details of the parameter estimates for the LNM-FA family are provided in the Appendix \ref{family}.

	\subsection{Initialization}
	For estimation, we need to first initialize the model parameters, variational parameters, and the component indicator variable $Z_{ig}$. The EM algorithm for finite mixture models is known to be heavily depending on starting values. Let  $z_{ig}^*$, $\pi_g^*$, $\boldsymbol{\mu}_g^*$, $\boldsymbol{D}_g^*$, $\boldsymbol{\Lambda}_{g}^*$,  $\m_{ig}^*$ and $\mV_{ig}^*$ be the initial values for $Z_{ig}$, $\pi_g$, $\boldsymbol{\mu}_g$, $\boldsymbol{D}_g$, $\boldsymbol{\Lambda}_{g}$,  $\m_{ig}$ and $\mV_{ig}$ respectively. The initialization is conducted as following: 
	\begin{enumerate}
\item $z_{ig}^*$ can be obtained by random allocation of observation to different clusters, cluster assignment from $k$-mean clustering or cluster assignment from some model-based clustering algorithms. Since our algorithm is based on a factor analyzer structure, we initialize $Z_{ig}$ using the cluster membership obtained by fitting parsimonious Gaussian mixture models\citep[PGMM;][]{McNicholas2008} to the transformed variable $\Y$ obtained using Equation \ref{eqn:Y}. For computational purposes, any 0 in the $\W$ were replaced by 0.001 for initialization. The implementation of PGMM is available in {\sf R} package ``pgmm"\citep{PGMM}. 
		\item Using this initial partition, $\boldsymbol{\mu}_g^*$ is the sample mean of the $g^{th}$ cluster and $\pi_g^*$ is the proportion of observations in the $g^{th}$ cluster in this initial partition.
		\vspace{0.1in}
		\item Similar to \cite{McNicholas2008}, we estimate the sample covariance matrix $\boldsymbol{S}_g^*$ for each group and then use eigendecomposition of  $\boldsymbol{S}_g^*$ to obtain $\boldsymbol{D}_g^*$ and $\boldsymbol{\Lambda}^*_g$. Suppose $\boldsymbol{\lambda}_g$ is a vector of the first $q$ largest eigenvalues of $\mathbf{S}_g^*$ and the columns of $\mathbf{L}_g$ are the corresponding eigenvectors, then
		\[
		 \boldsymbol{\Lambda}_{g}^*=\mathbf{L}_g\boldsymbol{\lambda}_g^{\frac{1}{2}}, \quad \text{and} \quad \boldsymbol{D}_g^*=\text{diag}\{\boldsymbol{S}_g^*-\boldsymbol{\Lambda}^*_g\boldsymbol{\Lambda}_g^{*^T}\}.
		\]

		\item As Newton-Raphson method is used to update the variational parameters, we need $\m^*$ and $\mV^*$. For $\m^*$, we apply an additive log ratio transformation on the observed taxa compositions $\hat{\p}$ and set $\m^*=\phi(\hat{\p})$ using Equation \ref{eqn:Y}. For $\mV^*$, we use a diagonal matrix where all diagonal entries are 0.1. Note that it is important to choose a small value for $\mV^*$ to avoid overshooting in Newton-Raphson method.  
	\end{enumerate}
	
	\subsection{Convergence, Model Selection and Performance assessment}
	Convergence of the algorithm is determined using Aitken acceleration criterion. The Aitken's acceleration \citep{aitken26} is defined as:
	\[
	a^{(k)}=\frac{l^{(k+1)}-l^{(k)}}{l^{(k)}-l^{(k-1)}}
	\]
	where $l^{(k+1)}$ stands for the log-likelihood values at $k+1$ iteration. Then, the asymptotic estimate for log-likelihood at iteration $k+1$ is:
	\[
	l_{\infty}^{(k+1)}=l^{(k)}+\frac{l^{(k+1)}-l^{(k)}}{1-a^{(k)}}
	\]
	The algorithm can be considered as converged when
	\[
	|l_{\infty}^{(k+1)}-l_{\infty}^{(k)}|<\epsilon
	\]
	where $\epsilon$ is a small number \citep{bohning94}. Here, we set $\epsilon=10^{-2}$.

	In clustering, the number of components are unknown. Hence, we run our algorithm for all possible numbers of clusters and latent variables, and the best model is chosen {\it a posteriori} using a model selection criteria. Here, we use the Bayesian Information Criterion \citep[BIC;][]{schwarz197801}. The BIC is the
	most popular choice for model selection in model-based clustering and is defined as	\[
	BIC=2l(\w,\boldsymbol{\hat{\vartheta}})-p\log(n),
	\]
	
	where $l(w,\boldsymbol{\hat{\vartheta}})$ is the log-likelihood evaluated using the estimated $\hat{\vartheta}$, $p$ is the number of free parameters, and $n$ is the
	number of observations. When the true labels are known, the Adjusted Rand Index \citep[ARI;][]{Hubert85comparingpartitions} is used for performance assessment. For perfect agreement, the ARI is 1 while the expected value of ARI is 0 under random classification.

	\section{Simulation Study}\label{sim}
		In this section, we use simulation studies to demonstrate the clustering performance and parameter recovery of the proposed LNM-FA models. We first generate $\mathbf{Y}$ from a multivariate normal distribution, then transform the data into composition $\boldsymbol{Z}$ using the additive log ratio transformation. Count data were then generated based on a multinomial distribution with composition $\boldsymbol{Z}$ with the total count for each observation being generated from a uniform distribution $U~[5000,10000]$. We conducted two sets of simulation studies, each comprising of 100 different datasets and we chose the best fitting model and the pair of $(G,q)$ using the BIC.		
	\subsection{Simulation Study 1}
\begin{figure}[!h]\centering\caption{Scatter plot of latent variable $\mathbf{Y}$ in one of the hundred datasets from Simulation Study 1. The observations are colored using their true class label. For this dataset, ARI of 1 was obtained by LNM-FA.}\label{fig:sim1}
			\includegraphics[width=5in,height=3.5in]{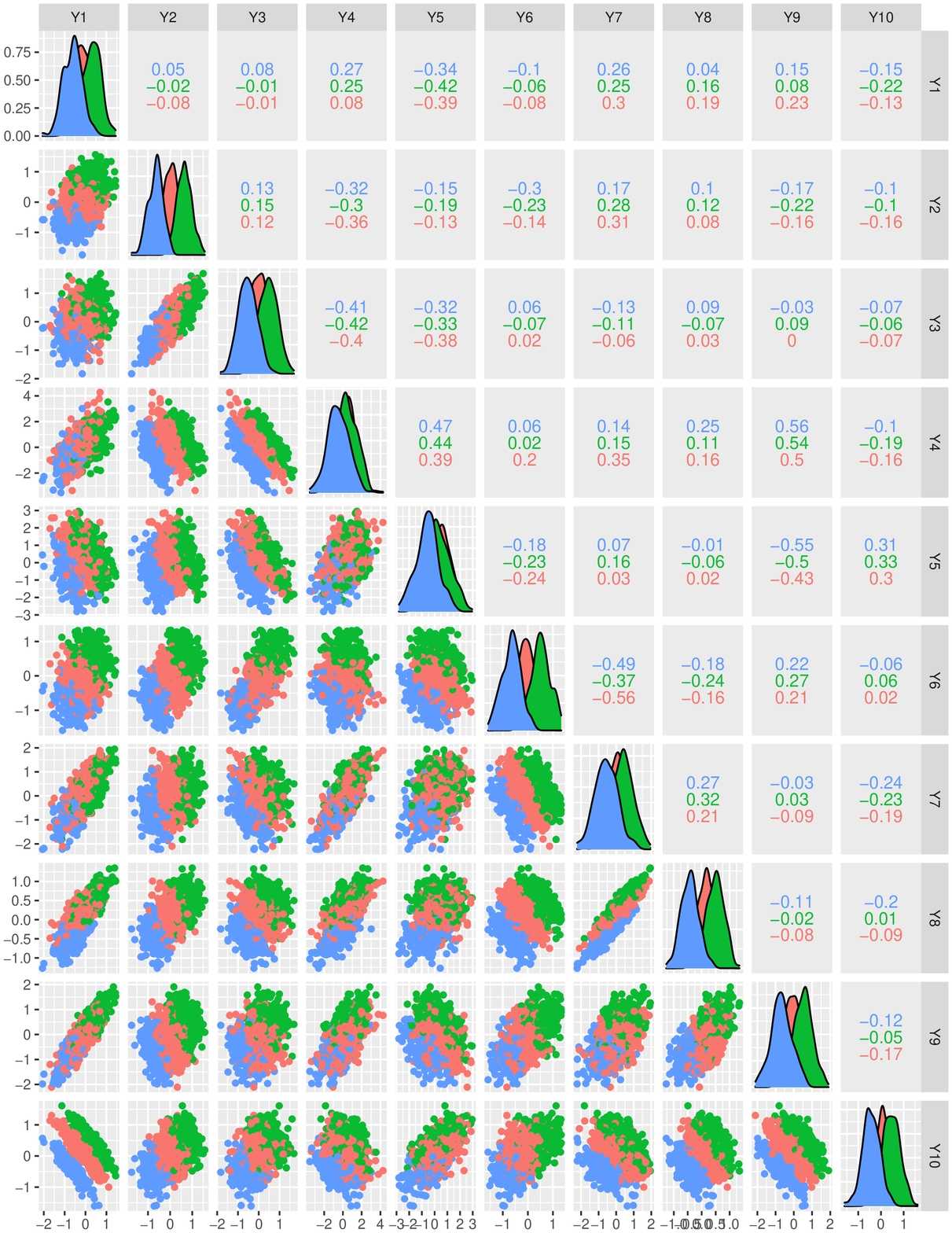}
		\end{figure}

\begin{figure}[!h]\centering\caption{Boxplot of the relative abundances of the observed counts in each cluster in one of the hundred datasets from Simulation Study 1.}\label{fig:sim1_relab}
			\includegraphics[height=\textwidth,angle=270]{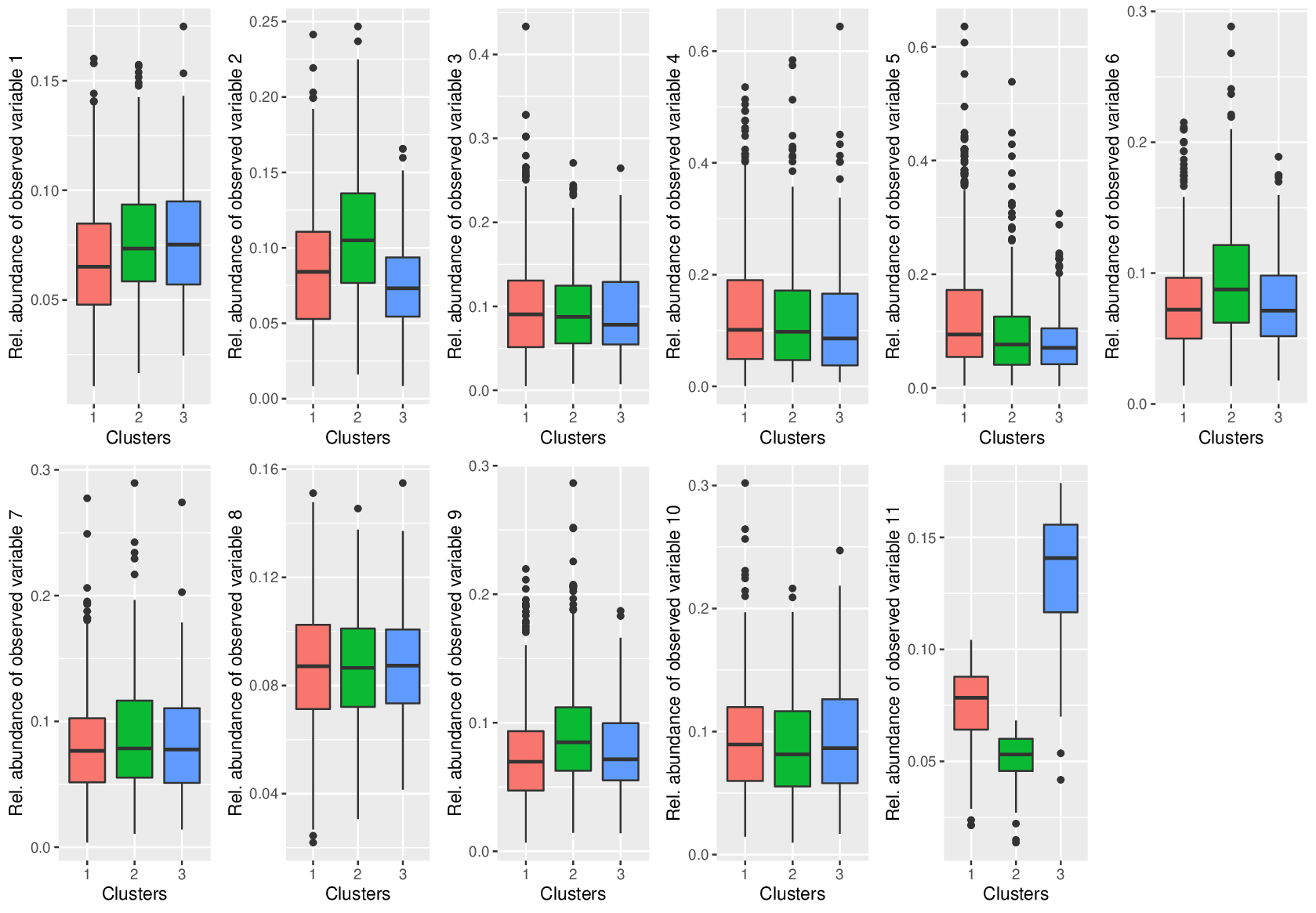}
		\end{figure}

	Here, we generated 100 ten-dimensional datasets, each of size $n = 1000$ from most constrained model ``CCC" with $G=3$, and $q=3$. Figure \ref{fig:sim1} shows a visualization of the cluster structure in the latent space for one of the hundred datasets and Figure \ref{fig:sim1_relab} shows the visualization of the relative abundance for observed count data of the same dataset. 
		
	We ran all 8 models in the LNM-FA family for $G = 1\ldots5$ and $q = 1\ldots5$ and selected the best model using the BIC.  In 96 out of 100 times, the BIC selected the true ``CCC" model with $G=3$ and $q=3$ with an average ARI of 0.999 (standard deviation [sd]  of 0.003). The true values of the parameters $\boldsymbol{\pi}$ and $\boldsymbol{\mu}$ are provided in Table \ref{tab:sim1}. As $\boldsymbol{\Lambda}$ is not identifiable but $\boldsymbol{\Sigma}=\boldsymbol{\Lambda}\boldsymbol{\Lambda}^T+\boldsymbol{D}$ is identifiable, we demonstrate the recovery of $\boldsymbol{\Sigma}$. The true value of $\boldsymbol{\Lambda}_g$ and $\boldsymbol{D}_g$ for $\boldsymbol{\Sigma}$ is provided in the Appendix \ref{sim:para} and the average and standard errors of norm of the bias of $\boldsymbol{\Sigma}$ is provided in Table \ref{tab:sim1}.

	\begin{table}[!htbp]
		\scriptsize
		\setlength{\tabcolsep}{5pt}
		\centering
		\caption{The true parameters and average of the estimated values along with the standard errors for Simulation Study 1.}\label{tab:sim1}
		\begin{tabular}{@{\extracolsep{10pt}}cc}
			\\[-1.8ex]\hline
			\hline \\[-1.8ex]
			\multicolumn{1}{c}{}& \multicolumn{1}{c}{Component 1($n_1=500$)}\\
			\hline
			&\\
			$\boldsymbol{\mu}_1$ &[-0.17,  0.03,  0.08,  0.24,  0.24, -0.06, -0.03,0.14, -0.11,  0.14]\\
			Average of $\hat{\boldsymbol{\mu}}_1$ &[-0.17,  0.03,  0.08,  0.25,  0.24, -0.06, -0.02,  0.14, -0.11,  0.14]\\
			sd of $\hat{\boldsymbol{\mu}}_1$ &(0.02, 0.01, 0.02 ,0.05, 0.04, 0.02, 0.03, 0.02, 0.03, 0.02)\\
			&\\
			$\pi_1$& 0.5\\
			Average of $\hat{\pi}_1$ (sd of $\hat{\pi}_1$)& 0.50(0.014)\\
			\hline \\[-1.8ex]
			\multicolumn{1}{c}{}& \multicolumn{1}{c}{Component 2($n_2=300$)}\\
			\hline 
			&\\
			$\boldsymbol{\mu}_2$ &[0.33, 0.63, 0.44, 0.60, 0.32, 0.52, 0.39, 0.50,0.51,0.45]\\
			Average of $\hat{\boldsymbol{\mu}}_2$ &[0.33,  0.63, ,0.44,  0.60,  0.33,  0.52,  0.39,  0.50,  0.51 , 0.45]\\
			sd of $\hat{\boldsymbol{\mu}}_2$&(0.03, 0.02, 0.03, 0.06, 0.05, 0.02, 0.04, 0.02, 0.03, 0.02)\\
			&\\
			$\pi_2$& 0.3\\
			Average of $\hat{\pi}_2$ (sd of $\hat{\pi}_2$)& 0.301(0.014)\\
			\hline \\[-1.8ex]
			\multicolumn{1}{c}{}& \multicolumn{1}{c}{Component 3($n_3=200$)}\\
			\hline
			&\\
			$\boldsymbol{\mu}_3$ &[-0.59, -0.66, -0.55, -0.45, -0.60, -0.68, -0.53, -0.41,-0.65, -0.46]\\
			Average of $\hat{\boldsymbol{\mu}}_3$ &[-0.587, -0.662, -0.553, -0.444 ,-0.602, -0.683, -0.526, -0.408, -0.647, -0.463 ]\\
			sd of $\hat{\boldsymbol{\mu}}_3$&(0.03, 0.03, 0.03, 0.07, 0.07, 0.03, 0.04, 0.02, 0.04, 0.03)\\
			&\\
			$\pi_3$& 0.2\\
			Average of $\hat{\pi}_3$ (sd of $\hat{\pi}_3$)& 0.199 (0.011)\\
			\hline \\[-1.8ex]
			\multicolumn{1}{c}{}& \multicolumn{1}{c}{Average and sd of the L1 norm of the difference between estimated and true Covariance.}\\
			\multicolumn{1}{c}{}& \multicolumn{1}{c}{Note that for ``CCC" model, all components have the same $\boldsymbol{\Sigma}$.}\\
			\hline
			&\\
			Average of $|\hat{\boldsymbol{\Sigma}}^{(i)}-\boldsymbol{\Sigma}|$&$0.85$\\
		&\\
			sd of $|\hat{\boldsymbol{\Sigma}}^{(i)}-\boldsymbol{\Sigma}|$&$0.27$\\
			\hline \\[-1.8ex]
		\end{tabular}
	\end{table}

For comparison, we also ran the LNM-MM and DMM on all hundred datasets for $G=1,\ldots,5$. In 81 out of the 100 datasets, the BIC selected a three component LNM-MM model with an average ARI of 0.99 (sd of 0.00) and a four component model in 13 of the datasets. The LNM-MM model encountered computational issues in 6 out of the 100 datasets. On the other hand, a five component DMM was selected by the BIC in all 100 datasets with an average ARI of 0.00 (sd 0.00).

	\subsection{Simulation Study 2}

\begin{figure}[!h]\centering\caption{Scatter plot of latent variable $\mathbf{Y}$ in one of the hundred datasets from Simulation Study 2. The observations are colored using their true class label. For this dataset, ARI of 1 was obtained by LNM-FA.}\label{fig:sim2}
			\includegraphics[width=5in,height=3.5in]{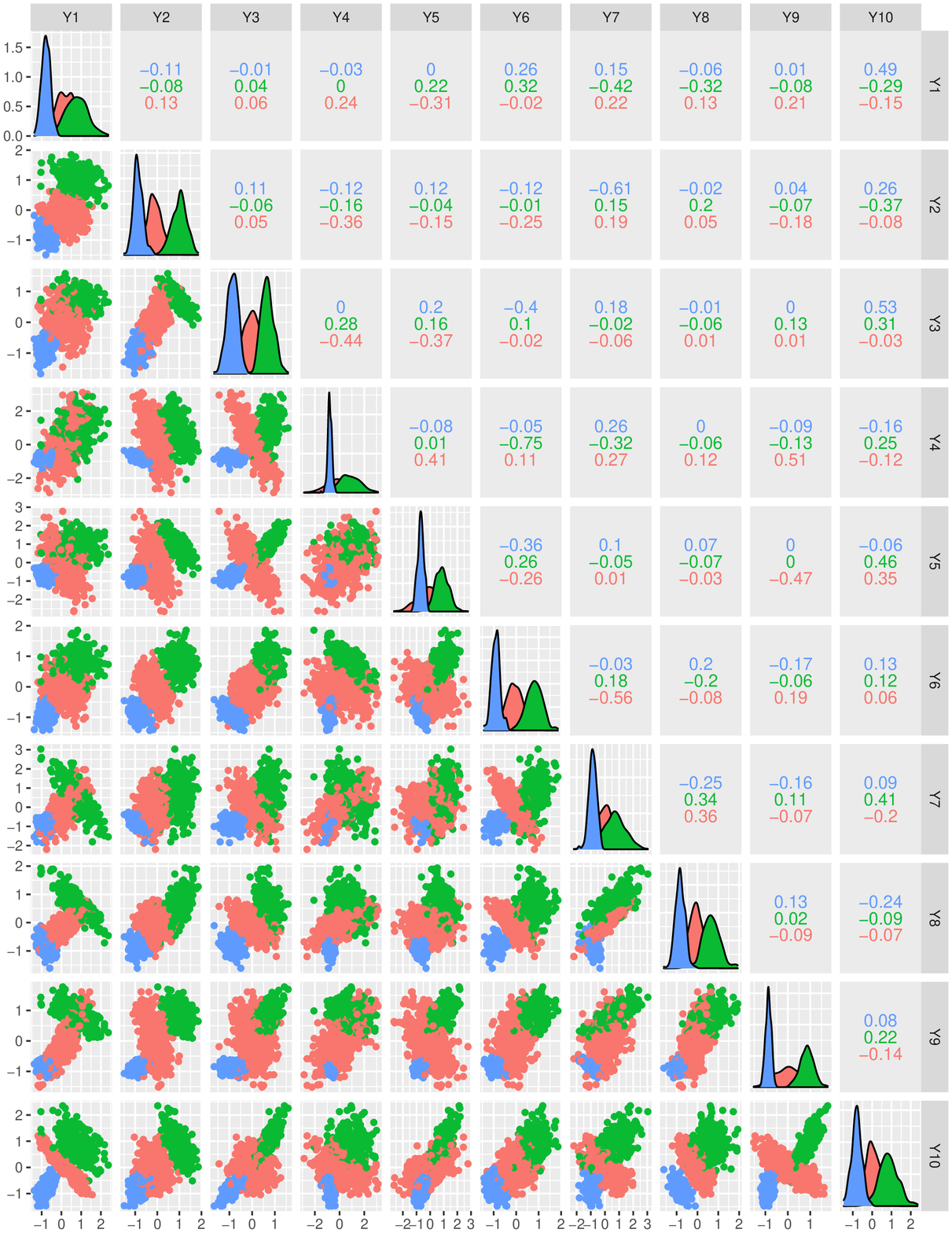}
		\end{figure}

\begin{figure}[!h]\centering\caption{Boxplot of the relative abundances of the observed counts in each cluster in one of the hundred datasets from Simulation Study 2.}\label{fig:sim2_relab}
			\includegraphics[height=\textwidth,angle=270]{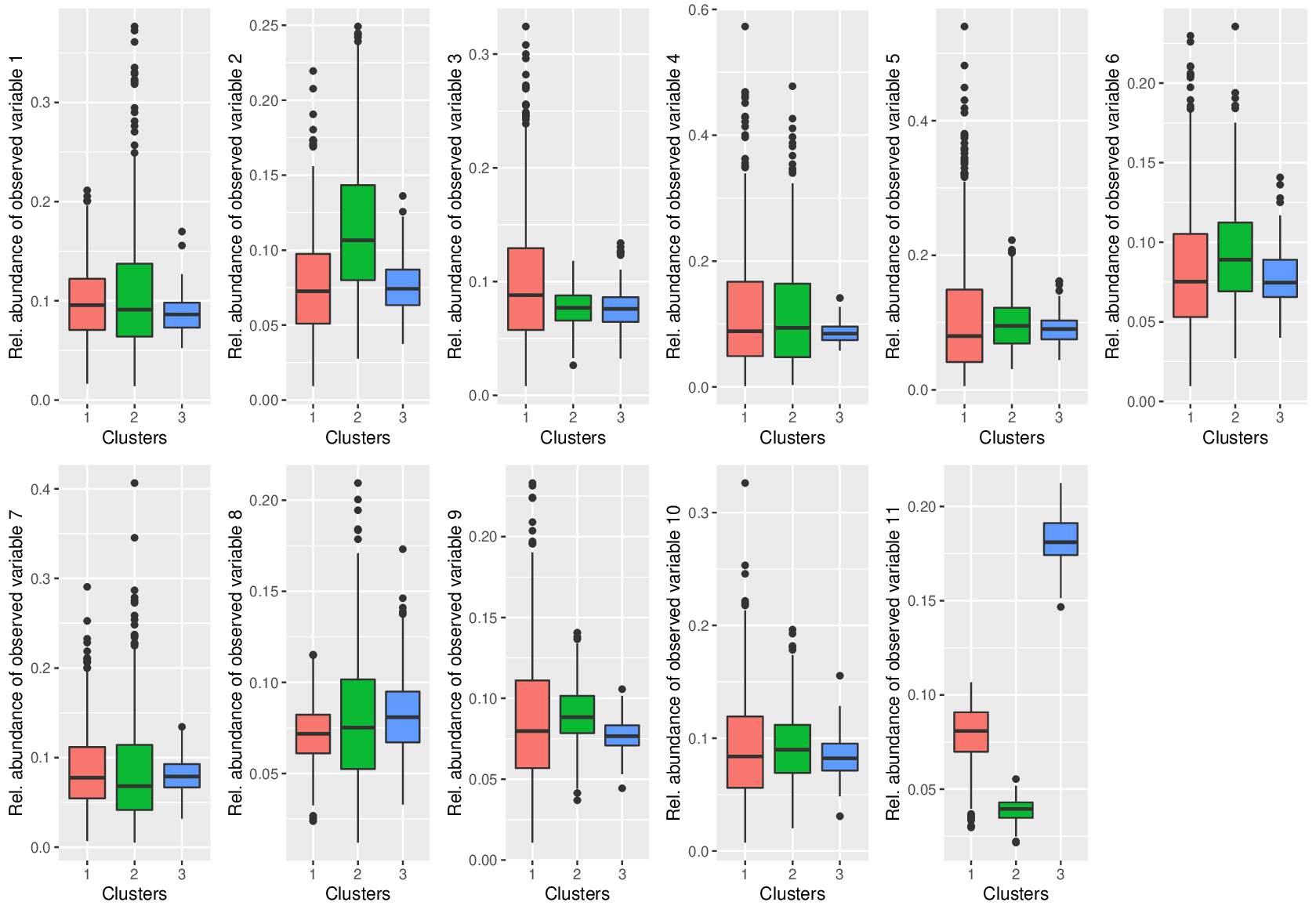}
		\end{figure}

	Here, we generate 100 ten-dimensional datasets, each of size $n = 1000$ from most flexible model ``UUU" with $G=3$, and $q=3$. Figure \ref{fig:sim2} shows visualization of the cluster structure in the latent space in one of the hundred datasets and Figure \ref{fig:sim2_relab} shows the visualization of the relative abundance for observed count data of the same dataset
	
	We ran all 8 models in the LNM-FA family for $G = 1\ldots5$ and $q = 1\ldots5$ and selected the best model using the BIC. In all 100 out of the 100 datasets, the BIC selected the true ``UUU" model with $G=3$ and $q=3$, with an average ARI of 1 (sd of 0). The true values of the parameters $\boldsymbol{\pi}$, $\boldsymbol{\mu}$ along with the average and standard deviations of their estimates are provided in Table \ref{tab:sim2}. Again, the average and standard deviations of the $L_1$ norm of $\boldsymbol{\Sigma}_g$ are provided in Table \ref{tab:sim2}.
	
	\begin{table}[!htbp]
		\scriptsize
		\setlength{\tabcolsep}{5pt}
		\centering
		\caption{The true parameters and average of the estimated values along with the standard errors for Simulation Study 2.}\label{tab:sim2}
		\begin{tabular}{@{\extracolsep{10pt}}cc}
			\\[-1.8ex]\hline
			\hline \\[-1.8ex]
		\multicolumn{1}{c}{}& \multicolumn{1}{c}{Component 1 ($n_1=500$)}\\
		\hline
		&\\
		
		$\boldsymbol{\mu}_1$ &[0.16, -0.13,  0.06,  0.13,  0.00, -0.06, -0.02, -0.11,  0.00,  0.03]\\
			Average of $\hat{\boldsymbol{\mu}}_1$ &[0.163 ,-0.130,  0.057,  0.134,  0.001, -0.064, -0.015, -0.108,  0.002,  0.027]\\
		sd of $\hat{\boldsymbol{\mu}}_1$  &(0.02, 0.01, 0.02, 0.05, 0.04, 0.02, 0.03, 0.02, 0.03, 0.02)\\[5pt]
		$\pi_1$& 0.5\\
		Average of $\hat{\pi}_1$ (sd of $\hat{\pi}_1$)& 0.50 (0.02)\\[5pt]
		Average of $|\hat{\boldsymbol{\Sigma}}_1^{(i)}-\boldsymbol{\Sigma}_1|$&$1.31$\\[5pt]
		sd of $|\hat{\boldsymbol{\Sigma}}_1^{(i)}-\boldsymbol{\Sigma}_1|$&$0.40$\\
		\hline \\[-1.8ex]
		\multicolumn{1}{c}{}& \multicolumn{1}{c}{Component 2 ($n_2=300$)}\\
		\hline 
		&\\
			$\boldsymbol{\mu}_2$ &[0.79, 1.01, 0.66, 0.76, 0.86, 0.83, 0.66, 0.68, 0.85, 0.84]\\
		Average of  $\hat{\boldsymbol{\mu}}_2$ &[0.79,  1.01, 0.66 , 0.76,  0.86,  0.83,  0.66,  0.68 , 0.85 , 0.84]\\
		sd of $\hat{\boldsymbol{\mu}}_2$ &(0.03, 0.02, 0.02, 0.05, 0.03, 0.03, 0.05, 0.02, 0.02, 0.03)\\
		&\\
		$\pi_2$& 0.3\\
		Average of  $\hat{\pi}_2$ (sd of $\hat{\pi}_2$) & 0.30(0.01)\\[5pt]
		Average of $|\hat{\boldsymbol{\Sigma}}_2^{(i)}-\boldsymbol{\Sigma}_2|$&$1.38$\\[5pt]
		SD of $|\hat{\boldsymbol{\Sigma}}_2^{(i)}-\boldsymbol{\Sigma}_2|$&$0.37$\\
		\hline \\[-1.8ex]
		\multicolumn{1}{c}{}& \multicolumn{1}{c}{Component 3 ($n_3=200$)}\\
		\hline
		&\\
			$\boldsymbol{\mu}_3$ &[-0.77, -0.89, -0.88, -0.78, -0.71, -0.89, -0.86, -0.82, -0.86, -0.80]\\
		Average of  $\hat{\boldsymbol{\mu}}_3$ &[-0.77, -0.89, -0.88, -0.780, -0.71, -0.89, -0.86, -0.82, -0.86, -0.80  ]\\
		sd of $\hat{\boldsymbol{\mu}}_3$ &(0.02, 0.02, 0.02, 0.01, 0.02, 0.01, 0.02, 0.01, 0.01, 0.02 )\\
		&\\
		$\pi_3$& 0.2\\
		Average of $\hat{\pi}_3$ (sd of $\hat{\pi}_3$)& 0.20 (0.01)\\[5pt]
		Average of $|\hat{\boldsymbol{\Sigma}}_3^{(i)}-\boldsymbol{\Sigma}_3|$&$0.38$\\[5pt]
		SD of $|\hat{\boldsymbol{\Sigma}}_3^{(i)}-\boldsymbol{\Sigma}_3|$&$0.06$\\	
			\hline \\[-1.8ex]
		\end{tabular}
	\end{table}

	We also ran the LNM-MM and DMM on all 100 datasets. From the LNM-MM family, the BIC selected a three component model in 12 out of the 100 datasets with perfect classification but four and five component models in 70 and 9 datasets, respectively. The LNM-MM implementation encountered singularities in the remaining 9 datasets. On the other hand, a five component DMM is selected every time with an average ARI of 0.27 (sd of 0.03).

	\section{Real data analysis}\label{real}
	We applied our method to three publicly available microbiome datasets.
	\begin{enumerate}[leftmargin=0em]
	\item[] \textbf{Dietswap Dataset:}
	We applied our algorithm to the microbiome dataset \texttt{Dietswap} \citep{O'Keefe} available in {\sf R} package \texttt{Microbiome} \citep{microbiome}. Colorectal cancer is the third most prevalent cancer worldwide \citep{garrett2019}. The rate of colon cancer in Americans of African descent is much higher than compared to rural Africans \citep{O'Keefe}. Recent findings indicate that the risk of colon cancer has been known to be associated with dietary habits that affects the gut microbiota \citep{garrett2019}. To investigate diet-associated cancer risk, \citep{O'Keefe} collected fecal samples from healthy middle aged 20 African(AFR) and 20 African American(AAM). Fecal samples were taken at 6 different timepoints:  the first three measurements (i.e., Day 0, Day 7, and Day 14) were taken in their home environment with their regular dietary habits and the last three measurements (i.e., Day 15, Day 22, and Day 29) were taken after an intervention diet. The Human Intestinal Tract Chip phylogenetic microarray was used for the global profiling of microbiota composition.
 As repeated measurements at different time points are taken on the same individuals and our model currently cannot model that (violates the independence assumption), we only utilize the measurements at Day 0. Hence, the resulting dataset comprises of 38 individuals from Day 0, and we focus our analysis at the genus level resulting in 130 genera. 
 
\item[] \textbf{FerrettiP Dataset:}
We applied our algorithm to the microbiome dataset \texttt{FerrettiP} \citep{ferretti} available in {\sf R} package \texttt{curatedMetagenomicData} \citep{curated}. The study sampled the microbiome of 25 mother-infant pairs  across multiple body sites from birth up to 4 months postpartum. Out of the 216 samples collected, 119 samples were derived from the stool (proxy for gut microbiome), 15 samples were derived from the skin swabs (skin microbiome), 63 samples were derived from the oral cavity swabs (oral microbiome), and 19 samples were derived from the vaginal swabs (vaginal microbiome). Here, we focus our analysis on the subset of the 119 stool samples (23 adults and 96 newborns).  As repeated measurements at different time points are taken on the same individuals (violates the independence assumption), we only focus on one time point (i.e., Day 1) for the newborns at the genus level. Hence, the resulting dataset comprises of 42 individuals (23 adults and 19 newborns) and 262 genera. 

\item[] \textbf{ShiB Dataset:} We applied our algorithm to the microbiome dataset \texttt{ShiB}\citep{ShiB} available in {\sf R} package \texttt{curatedMetagenomicData} \citep{curated}. Periodontitis is a common oral disease that affects about 50\% of the American adults and is associated with alterations in the subgingival microbiome of individual tooth sites \citep{ShiB}. Current commonly used clinical parameters cannot adequately predict the disease progression \citep{mcguire1996}. The study by \cite{ShiB} was designed to identify potential prognostic biomarkers using the compositions of the subgingival microbiome that can predict periodontitis. Oral microbiome samples were collected from 12 healthy individuals with chronic periodontitis  before and after nonsurgical therapy from multiples tooth sites.	Only the samples from the tooth sites that were clinically resolved after the therapy were retained, resulting in a total of 48 samples (24 periodontitis samples and 24 recovered samples) and 96 genera. Although multiple samples per individuals were obtained, \cite{ShiB} that individual tooth sites are likely to have independent clinical states and unique microbial communities in subgingival pockets, so we also treat samples as independent. 
\end{enumerate}

\begin{figure}\centering\caption{Heatmap of relative abundance of the differentially abundant genera in all three datasets.}\label{fig:heatmap}
\begin{subfigure}{.6\textwidth}
  \centering
	\includegraphics[width=0.9\textwidth,angle=270]{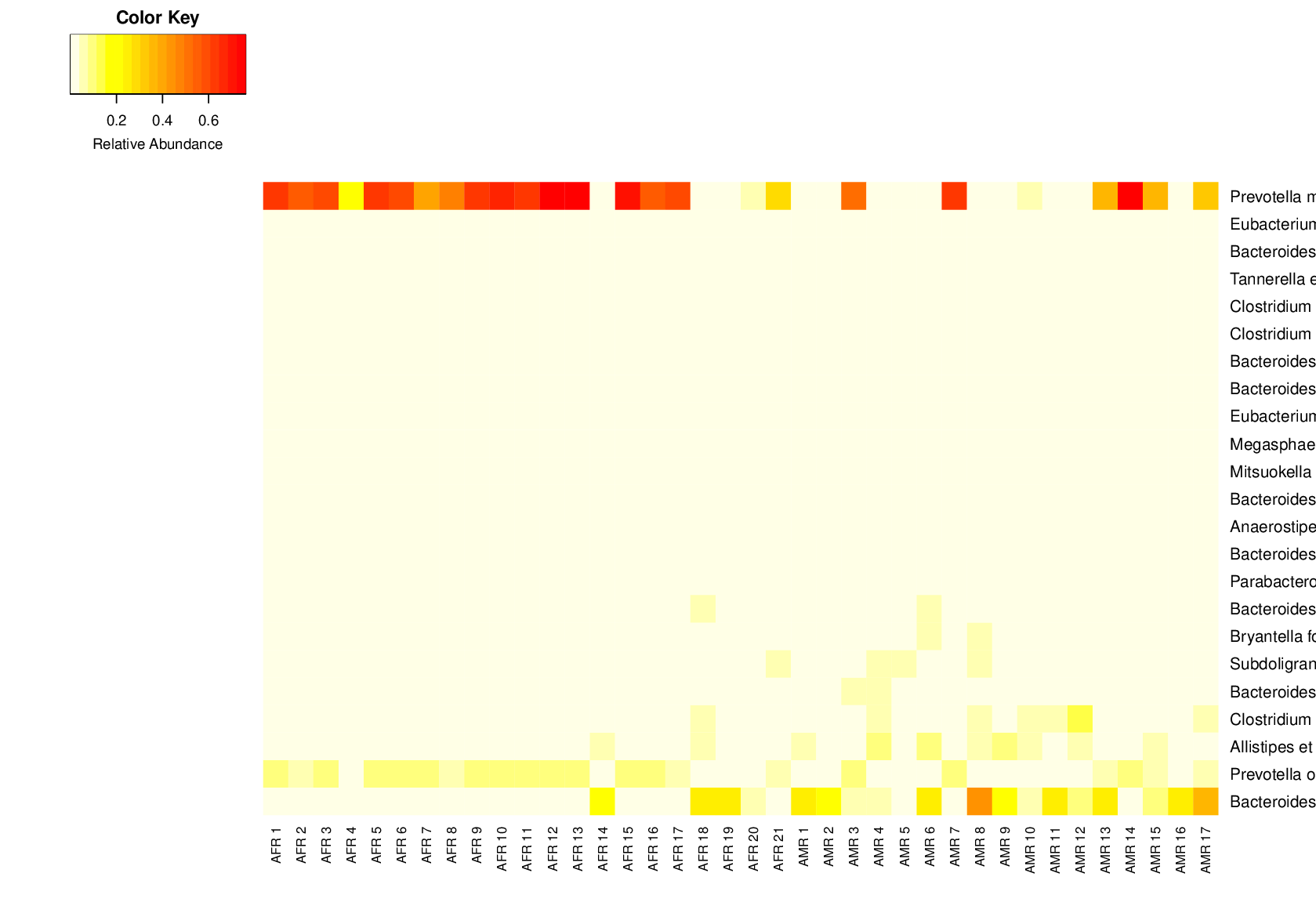}
  \caption{\texttt{Dietswap} dataset.}
    \end{subfigure}
    
  \begin{subfigure}{.49\textwidth}
  \centering
	\includegraphics[width=0.8\textwidth,angle=270]{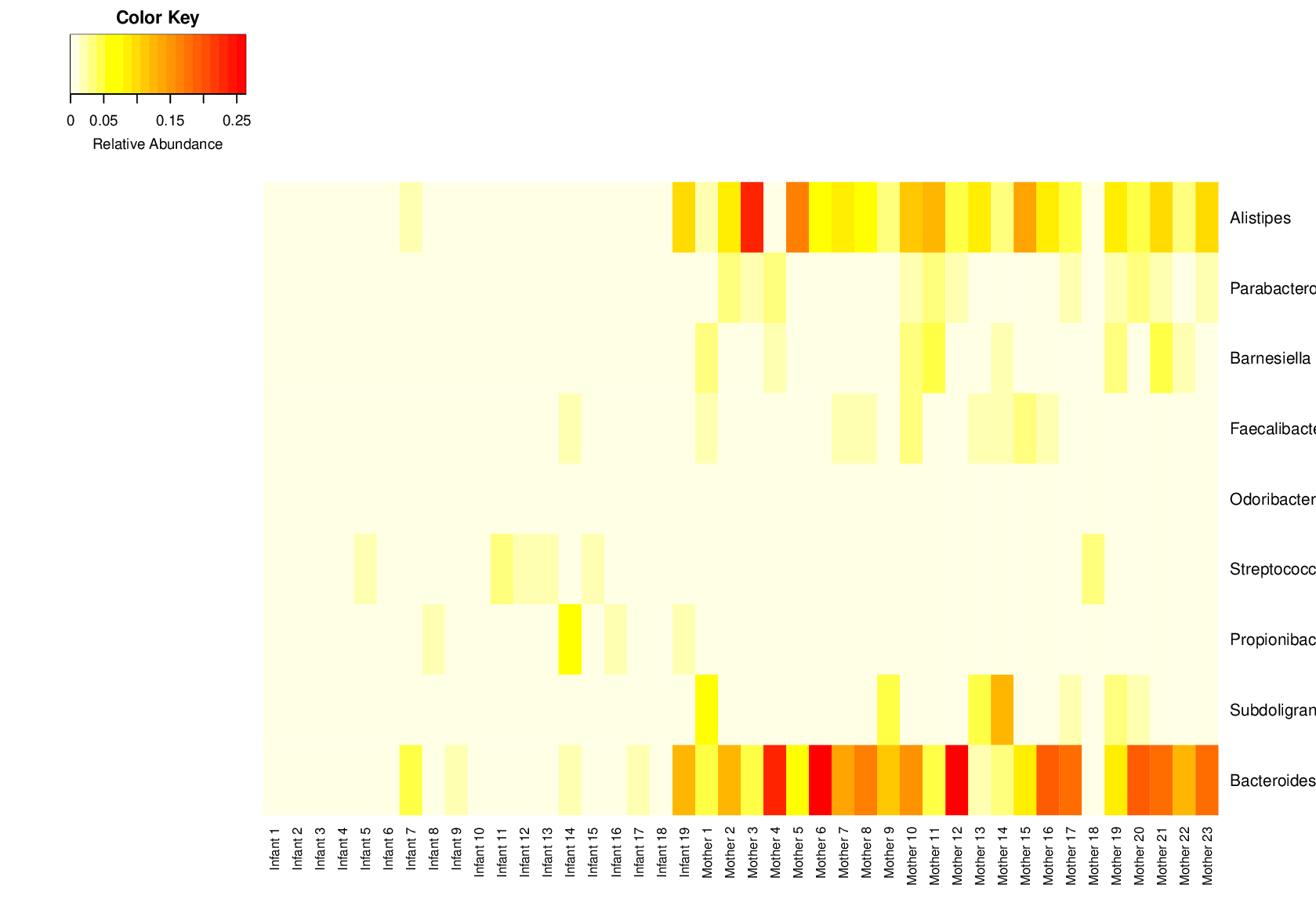}
  \caption{\texttt{FerrettiP} dataset.}
  \end{subfigure}
  \begin{subfigure}{.49\textwidth}
  \centering
	\includegraphics[width=0.8\textwidth,angle=270]{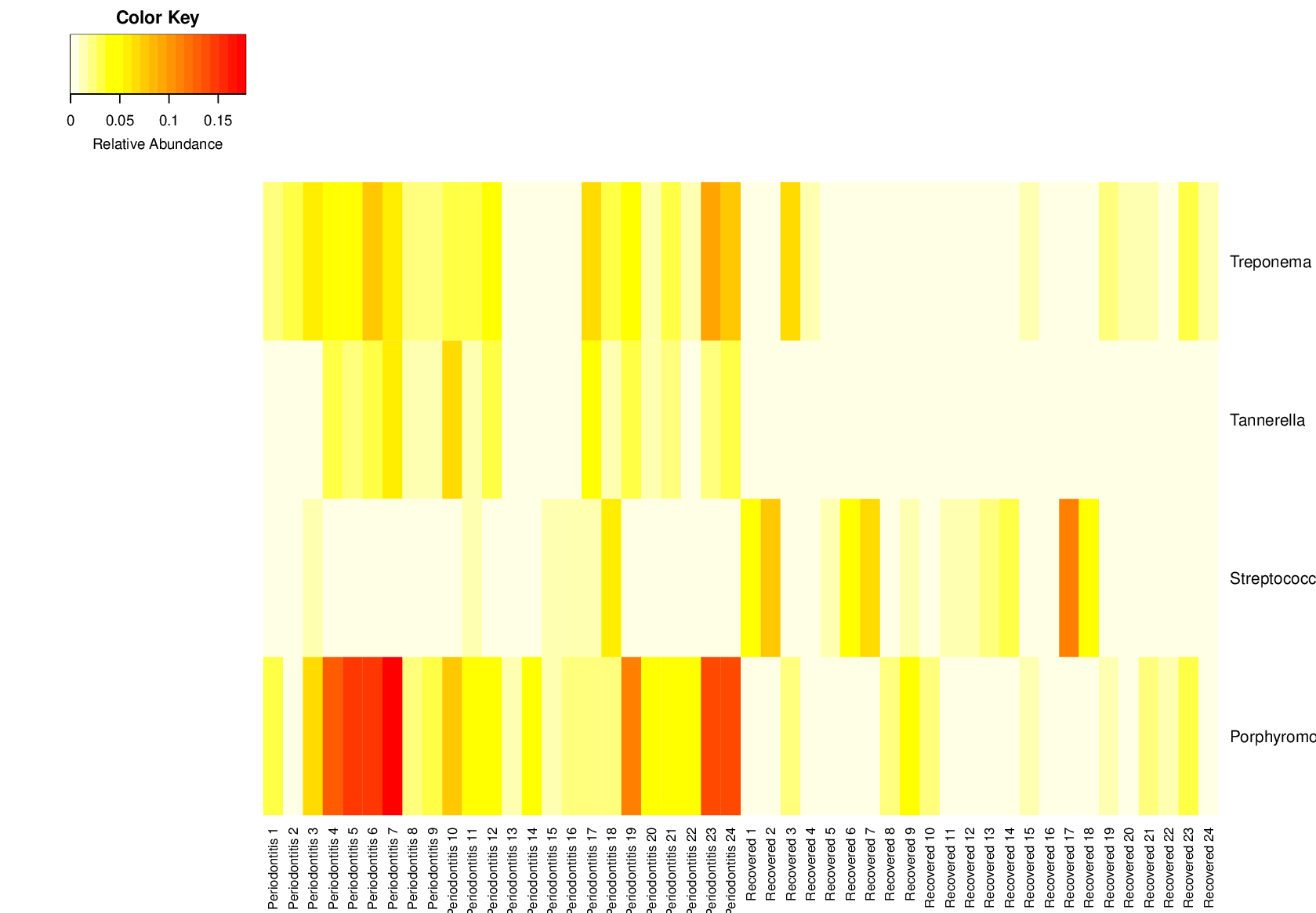}
  \caption{\texttt{ShiB} dataset.}
\end{subfigure}
\end{figure}

For all three datasets, we first utilize the {\sf R} package ALDEx2 \citep{ALDE1,ALDE2,ALDE3} for differential abundance analysis to identify the genera that are different among the two groups (i.e., AFR vs.\ AAM for \texttt{Dietswap} dataset, adults vs.\ infants for \texttt{FerrettiP} dataset, and periodontitis vs.\ recovered for \texttt{ShiB} dataset). This step is analogous to conducting differential expression analysis in RNA-seq studies before performing cluster analysis to remove the noise variables before clustering the data. Here, we used the Welch's $t$-test option in ALDEx2 on the log-transformed counts for each genera for differential abundance analysis and selected those genera for which the corresponding expected value of the Benjamini-Hochberg corrected p-value is less than 0.05. The numbers of differentially abundant genera for \texttt{Dietswap}, \texttt{FerrettiP}, and \texttt{ShiB} datasets are 23, 8, and 4, respectively. To preserve the relative abundance, the remaining genera are aggregated in a category ``Others", which is then used as the reference level for the additive log-ratio transformation.\\
The heatmaps of the relative abundances of the differentially expressed genera for all three datasets in Figure \ref{fig:heatmap} shows that there are some distinct differences in the relative abundance of the genera between the groups.
 
We ran all 8 models from our mixtures of LNM-FA family for $G = 1\ldots3$. Since all three datasets has different dimensions, we ran $q=1\ldots 5$ for \texttt{Dietswap} and \texttt{FerrettiP} datasets, and $q=1\ldots 3$ for \texttt{ShiB} dataset. The BIC was used to select the best fitting model. For comparison, we also ran the mixtures of LNM models without the factor structure and the Dirichlet-multinomial mixture model on all three datasets for $G = 1 \ldots 3$. The classification results from all three approaches are summarized in Table \ref{tab:realdata}

	\begin{table}[!htbp]
		\centering
		\caption{Summary of the clustering performances on all three real datasets using best fitting model by LNM-FA, LNM-MM and DMM.}\label{tab:realdata}
		\begin{tabular}{@{\extracolsep{\fill}}ccccc|ccc}
			\\[-1.8ex]\hline
			\hline \\[-1.8ex]
			
			Data&Approach&\multicolumn{2}{c}{Estimated} &\multicolumn{3}{c}{Classification Table}& ARI\\
			&(Model)&G&q&\multicolumn{4}{c}{~}\\
			\hline
			&&&  &&\textbf{AAM}&\textbf{AFR}&\\\cline{5-7}
			&LNM-FA&2& 2 &Est. Group 1&20&1&\textbf{0.8}\\
			&(CUU)&&  &Est. Group&1&16&\\[5pt]\cline{2-8}\\[-10pt]
			&&&  &&\textbf{AAM}&\textbf{AFR}&\\\cline{5-7}
			\texttt{Dietswap}&LNM-MM&-& - &-&-&-&-\\
			&&& &-&-&-&\\[5pt]\cline{2-8}\\[-10pt]
			&&&  &&\textbf{AAM}&\textbf{AFR}&\\\cline{5-7}
			&DMM&3& - &Est. Group 1&2&13&0.38\\
			&&& &Est. Group 2&14&1&\\
			&&& &Est. Group 3&5&3&\\
			\\[-1.8ex]\hline
			\hline
			&&&  &&\textbf{Adult}&\textbf{Infant}&\\\cline{5-7}
			&LNM-FA&2& 1 &Est. Group 1&22&0&\textbf{0.9}\\
			&(UCC)&&  &Est. Group 2&1&19&\\[5pt]\cline{2-8}\\[-10pt]
			&&&  &&\textbf{Adult}&\textbf{Infant}&\\\cline{5-7}
			\texttt{FerrettiP}&LNM-MM&-& - &-&-&-&-\\
			&&& &-&-&-&\\[5pt]\cline{2-8}\\[-10pt]
			&&&  &&\textbf{Adult}&\textbf{Infant}&\\\cline{5-7}
			&DMM&2& - &Est. Group 1&22&1&0.81\\
			&&& &Est. Group 2&1&18&\\
			\\[-1.8ex]\hline
			\hline
			&&&  &&\textbf{Periodontitis}&\textbf{Recovered}&\\\cline{5-7}
			&LNM-FA&2& 1 &Est. Group 1&21&4&\textbf{0.49}\\
			&(CCC)&&  &Est. Group 2&3&20&\\[5pt]\cline{2-8}\\[-10pt]
			&&&  &&\textbf{Periodontitis}&\textbf{Recovered}&\\\cline{5-7}
			\texttt{ShiB}&LNM-MM&2& - &Est. Group 1&21&4&\textbf{0.49}\\
			&&& &Est. Group 2&3&20&\\[5pt]\cline{2-8}\\[-10pt]
			&&&  &&\textbf{Periodontitis}&\textbf{Recovered}&\\\cline{5-7}
			&DMM&2& - &Est. Group 1&18&2&0.43\\
			&&& &Est. Group 2&6&22&\\
			\\[-1.8ex]\hline
			\hline
		\end{tabular}
	\end{table}

In all three datasets, our proposed LNM-FA was able to recover the underlying groups. Our proposed approach outperformed DMM in all three datasets. In the \texttt{Dietswap} (sample size $n=38$) and \texttt{FerrettiP} datasets (sample size $n=42$), 23 and 8 taxa in genus level were identified as  differentially abundant, respectively. Thus, while fitting LNM-MM model in these two datasets, $\boldsymbol{\Sigma}$ becomes singular, while the LNM-FA could be fitted due to the computational advantage that comes with the incorporation of factor analyzer structure. On the other hand, the LNM-MM could be fitted for \texttt{ShiB} dataset where the dimensionality of the dataset after differential abundance analysis was 5 (i.e., four differentially abundant genera and one aggregated column of ``Others"). In \texttt{ShiB} dataset, both LNM-FA and LNM-MM selected a two component model with an ARI of 0.49. However, the number of parameters that needs to be estimated for the covariance matrices of the latent variable in best fitting model by LNM-FA (i.e., CCC with q=1) is less compared to the LNM-MM (i.e., 4 for LNM-FA vs.\ 20 for LNM-MM). Note, that the DMM model could be fitted to all three datasets. The DMM model accounts for overdispersion by utilizing a Dirichlet prior on the multinomial parameter $\mathbf{p}$. However, as noted by \cite{Aitchison} and \cite{Xia}, the logistic normal multinomial distribution allows for a more flexible covariance structure than the Dirichlet-multinomial model.

	\section{Conclusion}\label{conc}
	Here,  we extended the additive logistic normal multinomial mixture model for high dimensional data by incorporating a factor analyzer structure. A family of eight mixture models was proposed by imposing constraints on the components of the covariance matrix of the latent variable. Due to the incorporation of the factor analyzer structure, the number of parameters are now linear in the dimensionality of the latent variable as opposed to the additive logistic normal multinomial mixture model where the number of parameters grows quadratically. Through simulation studies, we demonstrated that our proposed approach provides  excellent clustering performance and parameter recovery. Imposing a factor analyzer structure allows us to work on a lower dimension $q$ compare to $K$ and thus, the number of free parameters in the covariance matrix is greatly reduced when $q$ is chosen to be sufficiently smaller than $K$. Additionally, the use of Woodbury identity provides additional computational advantages. For the real data analysis, our approach outperforms the Dirichlet-multinomial mixture model in all three datasets. For the \texttt{Dietswap} dataset and \texttt{FerrettiP} datasets, the LNM-MM by \citep{fang2020}(i.e., the additive logistic multinomial mixture model without the utilize factor analyzer structure) could not be fitted due to computational issues as the dimensions of those datasets are higher. In \texttt{ShiB} dataset where $K$ is small, the LNM-MM and our proposed LNM-FA provide comparable performance. While our approach can deal with high dimensional nature of the microbiome data, it does not account for any covariate information currently. Microbiome composition is very dynamic and is affected by time variant covariates such as diet, environmental exposures and time invariant covariates such as gender. Understanding how various biological/environmental factors affect the changes in the microbiome compositions might be valuable in gaining valuable biological insight into disease diagnosis and prognosis.

	\section*{Acknowledgements}
This work was supported by the Collaboration Grant for Mathematicians from the Simons Foundation.

	\appendix
	\section{ELBO for LNM model}\label{ELBO}
First, we decompose $F(q(\y),\w)$ into 3 parts:
	\[
	F(q(\y),\w)=\int q(\boldsymbol{y})\log f(\boldsymbol{w}|\boldsymbol{y}) d\boldsymbol{y} + \int q(\boldsymbol{y})\log f(\boldsymbol{y}) d\boldsymbol{y} - \int q(\boldsymbol{y})\log q(\boldsymbol{y}) d\boldsymbol{y}.
	\]
	The second and third integral (i.e. $E_{q(\y)}(\log f(\y))$ and $E_{q(\y)}(\log q(\y))$) have explicit solutions such that
	\[
	E_{q(\y)}(\log f(\y))=-\dfrac{K}{2}\log(2\pi)-\frac{1}{2}\log|\boldsymbol{\Sigma}|-\frac{1}{2}(\boldsymbol{m}-\boldsymbol{\mu})^T\boldsymbol{\Sigma}^{-1}(\boldsymbol{m}-\boldsymbol{\mu})-\frac{1}{2} \text{tr}(\boldsymbol{\Sigma}^{-1}\boldsymbol{\mV})
	\]
	and
	\[
	-E_{q(\y)}(\log q(\y))=\frac{1}{2}\log|\mV|+\dfrac{K}{2}+\frac{K}{2}\log(2\pi).
	\]
	Note that $\mV$ is a diagonal matrix. 
	As for the first integral, it has no explicit solution because of the expectation of log sum exponential term:
	\[
	E_{q(\y)}(\log f(\y|\w))=C+{\w^*}^T\m-\left(\sum_{k=1}^{K+1}\w_k\right)E_{q(\y)}\left[\log\sum_{k=1}^{K+1}\exp \y_k\right]
	\]
	where $\w^*$ represents a $K$ dimension vector with first $K$ elements of $\w$, $\y_{K+1}$ is set to 0 and $C$ stands for $\log\frac{\boldsymbol{1}^T\w!}{\prod_{k=1}^{K}\w_{k}!}$. \cite{Blei07acorrelated} proposed an upper bound for $E_{q(\y)}\left[\log\left(\sum_{k=1}^{K+1}\exp \y_k\right)\right]$ as
\begin{equation}
E_{q(\y|\m,\mV)}\left[\log\left(\sum_{k=1}^{K+1}{\exp\y_k}\right)\right] \leq \xi^{-1}\left\{\sum_{k=1}^{K+1}E_{q(\mathbf{\y}|\m,\mV)}\left[\exp(\y_k)\right]\right\}-1+\log(\xi),\label{Blei}
\end{equation}
where $\xi\in \real$ is introduced as a new variational parameter. \cite{fang2020} utilized this upper bound to find a lower bound for $E_{q(\y)}(\log f(\y|\w))$. Here we further simplify lower bound by \cite{Blei07acorrelated}. Let $\mathbf{Z}=\sum_{k=1}^{K+1}\exp(\y_k)$, then we have:
	\begin{equation*}
	\begin{split}
	&E_{q(\y)}\left[\log\left(\sum_{k=1}^{K+1}\exp \y_k\right)\right] \leq \log E_{q(\y)}\left(\sum_{k=1}^{K+1}\exp \y_k \right) = \log\left[\sum_{k=1}^{K}\exp\left(\m_k+\frac{\mV_k}{2}\right)+1\right],
	\end{split}
	\end{equation*}
	where $\m_k, \mV_k$ stands for $k^{th}$ entry of $\m$ and the $k^{th}$ diagonal entry of $\mV$. The two upper bounds are equal when minimize \ref{Blei} with respect to $\xi$. . 
	
	Combining all 3 parts together, we have the approximate lower bound for $\log f(\w)$:
	\begin{equation*}
	\begin{split}
	\tilde{F}(q(\y),\w)&=C+{\w^*}^T\m-\left(\sum_{k=1}^{K+1}\w_k\right)\left\{\log\left[\sum_{k=1}^{K}\exp\left(\m_k+\frac{\mV_k}{2}\right)+1\right]\right\}+\\
	&\frac{1}{2}\log|\mV|+\dfrac{K}{2}-\frac{1}{2}\log|\boldsymbol{\Sigma}|-\frac{1}{2}(\boldsymbol{m}-\boldsymbol{\mu})^T\boldsymbol{\Sigma}^{-1}(\boldsymbol{m}-\boldsymbol{\mu})-\frac{1}{2} \text{tr}(\boldsymbol{\Sigma}^{-1}\boldsymbol{\mV})
	\end{split}
	\end{equation*}

	\section{ELBO for Cycle 2}\label{cycle2}
Here, in the second cycle, we have 
\begin{align*}
	F(q(\u,\y),\w)&=\int q(\u,\y)\log \frac{f(\w,\u,\y)}{q(\u,\y)}d\y d\u\\
	&=\int q(\u,\y)\log f(\w|\u,\y)d\y d\u+\int q(\u,\y)\log f(\u,\y)d\y d\u\\
	&-\int q(\u,\y)\log q(\u,\y)d\y d\u.
\end{align*}

	Furthermore, we assume that $q(\u,\y)=q(\u)q(\y)$, $\u\sim N(\tilde{\m},\tilde{\mV})$ and $\y \sim N(\m,\mV)$. Thus, the first term can be written as:
\begin{align*}
	\int q(\u,\y)\log f(\w|\u,\y)d\y d\u &= \int q(\u)q(\y)\log f(\w|\y) d\y d\u \\
	&= \int q(\y)\log f(\w|\y) d\y 
	\end{align*}
	This is identical to the first term in the ELBO in the first cycle and thus its lower bound is
	\[
	\int q(\u,\y)\log f(\w|\u,\y)d\y d\u \geq C+{\w^*}^T\m-\left(\sum_{k=1}^{K+1}\w_k\right)\left\{\log\left(\sum_{k=1}^{K}\exp\left(\m_k+\frac{\mV_k}{2}\right)+1\right)\right\}
	\]
	
	The third term is
	\[
	-\int q(\u,\y)\log q(\u,\y)d\y d\u=\frac{1}{2}\left(\log |\mV|+\log|\tilde{\mV}|+q+K+(K+q)\log 2\pi \right).
	\]

	The second term is
	\begin{equation*}
	\begin{split}
	\int q(\u,\y)\log f(\u,\y)d\y d\u&=\int q(\u)q(\y)\log[f(\y|\u)f(\u)]d\y d\u\\
	=&~E_{q(\u)}E_{q(\y)}(\log f(\y|\u)f(\u))\\
	=&-\frac{1}{2}\left\{(q+K)\log(2\pi)-\log|\boldsymbol{D}|-\tilde{\m}^T\tilde{\m}-\text{tr}(\tilde{\mV})-\text{tr}(\boldsymbol{\Lambda}^T\boldsymbol{D}^{-1}\boldsymbol{\Lambda}\tilde{\mV})\right.\\
	&-\text{tr}\left(\boldsymbol{D}^{-1}(\boldsymbol{V}+(\boldsymbol{m}-\boldsymbol{\mu})^T(\boldsymbol{m}-\boldsymbol{\mu}))\right)+2(\boldsymbol{m}-\boldsymbol{\mu})^T\boldsymbol{D}^{-1}\boldsymbol{\Lambda}\tilde{\m}\\
	&\left.-\tilde{\m}^T\boldsymbol{\Lambda}^T\boldsymbol{D}^{-1}\boldsymbol{\Lambda}\tilde{\m}\right\}.
	\end{split}
	\end{equation*}

	Overall, the ELBO in second cycle is:
	\begin{equation*}
	\begin{split}
	F(q(\u,\y),\w)&\geq C+\w^T\m-\left(\sum_{i=1}^{K+1}\w_i\right)\{\log(\sum_{k=1}^{K}\exp(\m_k+\frac{\mV_k}{2})+1)\}+\\
	&\frac{1}{2}(\log |\mV|+\log|\tilde{\mV}|+q+K-\log|\boldsymbol{D}|-\tilde{\m}^T\tilde{\m}-tr(\tilde{\mV})-\\
	&tr(\boldsymbol{D}^{-1}(\boldsymbol{V}+(\boldsymbol{m}-\boldsymbol{\mu})^T(\boldsymbol{m}-\boldsymbol{\mu})))+2(\boldsymbol{m}-\boldsymbol{\mu})^T\boldsymbol{D}^{-1}\boldsymbol{\Lambda}\tilde{\m}-\\
	&\tilde{\m}^T\boldsymbol{\Lambda}^T\boldsymbol{D}^{-1}\boldsymbol{\Lambda}\tilde{\m}-tr(\boldsymbol{\Lambda}^T\boldsymbol{D}^{-1}\boldsymbol{\Lambda}\tilde{\mV}))
	\end{split}
	\end{equation*}
	where $\m$ and $\mV$ are calculated from first stage. \\
	
	In addition to variational parameter in second stage, it is worth to notice that $\tilde{\m}_{ig}=E(\u_{ig}|\y_i, z_{ig})$, and $\tilde{\mV}_{g}=Cov(\u_{ig}|\y_i, z_{ig})$. Because the following relationship:  
	\[
	\left[\begin{matrix}
	\boldsymbol{y}_i\\
	\boldsymbol{u}_{ig}
	\end{matrix}
	\right]|z_{ig}\sim MVN\left[\begin{matrix}
	\left(\begin{matrix}
	\boldsymbol{\mu}_g\\
	0
	\end{matrix}\right), & \left(\begin{matrix}
	\boldsymbol{\Lambda}_g\boldsymbol{\Lambda}_g^T+\boldsymbol{D}_g & \boldsymbol{\Lambda}_g\\
	\boldsymbol{\Lambda}_g^T & \mathbf{I}_q
	\end{matrix}\right)
	\end{matrix}\right]
	\]
	Therefore: 
	\[
	E(\boldsymbol{u}_{ig}|\boldsymbol{y}_i, z_{ig}=1)=\boldsymbol{\Lambda}_g^T(\boldsymbol{\Lambda}_g\boldsymbol{\Lambda}_g^T+\boldsymbol{D}_g)^{-1}(\boldsymbol{m}_{ig}-\boldsymbol{\mu}_g)
	\]
	\[
	Cov(\u_{ig}|\y_i, z_{ig})=\mathbf{I}_q-\boldsymbol{\Lambda}_g^T(\boldsymbol{\Lambda}_g\boldsymbol{\Lambda}_g^T+\boldsymbol{D}_g)^{-1}\boldsymbol{\Lambda}_g
	\]
	Then because the following inverse can be split as:
	\[
	(\boldsymbol{\Lambda}_g^T\boldsymbol{D}_{g}^{-1}\boldsymbol{\Lambda}_g+\mathbf{I}_q)^{-1}=\mathbf{I}_q-\boldsymbol{\Lambda}_g^T\boldsymbol{D}_{g}^{-\frac{1}{2}}(\boldsymbol{I}+\boldsymbol{D}_{g}^{-\frac{1}{2}}\boldsymbol{\Lambda}_g\boldsymbol{\Lambda}_g^T\boldsymbol{D}_{g}^{-\frac{1}{2}})^{-1}\boldsymbol{D}_{g}^{-\frac{1}{2}}\boldsymbol{\Lambda}_g
	\]
	and because $\boldsymbol{D}_g$ is always invertible by design. we have:
	\[
	\tilde{\mV}=(\boldsymbol{\Lambda}_g^T\boldsymbol{D}_{g}^{-1}\boldsymbol{\Lambda}_g+\mathbf{I}_q)^{-1}=\mathbf{I}_q-\boldsymbol{\Lambda}_g^T(\boldsymbol{D}_g+\boldsymbol{\Lambda}_g\boldsymbol{\Lambda}_g^T)^{-1}\boldsymbol{\Lambda}_g
	\]
	Above shows $\tilde{\mV}=Cov(\u_{ig}|\y_i, z_{ig})$. Apply same trick on $\tilde{\m}$, we have:
	\begin{equation*}
	\begin{split}
	\tilde{\m}&=(\boldsymbol{\Lambda}_g^T\boldsymbol{D}_{g}^{-1}\boldsymbol{\Lambda}_g+\mathbf{I}_q)^{-1}\boldsymbol{\Lambda}_g^T\boldsymbol{D}_g^{-1}(\m_{ig}-\boldsymbol{\mu}_g)\\	
	&=(\mathbf{I}_q-\boldsymbol{\Lambda}_g^T(\boldsymbol{D}_g+\boldsymbol{\Lambda}_g\boldsymbol{\Lambda}_g^T)^{-1}\boldsymbol{\Lambda}_g)\boldsymbol{\Lambda}_g^T\boldsymbol{D}_g^{-1}(\m_{ig}-\boldsymbol{\mu}_g)\\
	&=\boldsymbol{\Lambda}_g^T(\boldsymbol{D}_g^{-1}-(\boldsymbol{D}_g+\boldsymbol{\Lambda}_g\boldsymbol{\Lambda}_g^T)^{-1}\boldsymbol{\Lambda}_g\boldsymbol{\Lambda}_g^T\boldsymbol{D}_g^{-1})(\m_{ig}-\boldsymbol{\mu}_g)\\
	&=\boldsymbol{\Lambda}_g^T(\boldsymbol{\Lambda}_g\boldsymbol{\Lambda}_g^T+\boldsymbol{D}_g)^{-1}(\boldsymbol{m}_{ig}-\boldsymbol{\mu}_g)
	\end{split}
	\end{equation*}
	Where the last equality is followed by: 
	\begin{equation*}
	\begin{split}
	\boldsymbol{I}&=(\boldsymbol{\Lambda}_g\boldsymbol{\Lambda}_g^T+\boldsymbol{D}_g)(\boldsymbol{D}_g^{-1}-(\boldsymbol{D}_g+\boldsymbol{\Lambda}_g\boldsymbol{\Lambda}_g^T)^{-1}\boldsymbol{\Lambda}_g\boldsymbol{\Lambda}_g^T\boldsymbol{D}_g^{-1})\\
	&=(\boldsymbol{D}_g^{-1}-(\boldsymbol{D}_g+\boldsymbol{\Lambda}_g\boldsymbol{\Lambda}_g^T)^{-1}\boldsymbol{\Lambda}_g\boldsymbol{\Lambda}_g^T\boldsymbol{D}_g^{-1})^T(\boldsymbol{\Lambda}_g\boldsymbol{\Lambda}_g^T+\boldsymbol{D}_g)^T\\
	&=(\boldsymbol{D}_g^{-1}-(\boldsymbol{D}_g+\boldsymbol{\Lambda}_g\boldsymbol{\Lambda}_g^T)^{-1}\boldsymbol{\Lambda}_g\boldsymbol{\Lambda}_g^T\boldsymbol{D}_g^{-1})(\boldsymbol{\Lambda}_g\boldsymbol{\Lambda}_g^T+\boldsymbol{D}_g)
	\end{split}
	\end{equation*}
	Because of symmetric. We showed that $(\boldsymbol{D}_g^{-1}-(\boldsymbol{D}_g+\boldsymbol{\Lambda}_g\boldsymbol{\Lambda}_g^T)^{-1}\boldsymbol{\Lambda}_g\boldsymbol{\Lambda}_g^T\boldsymbol{D}_g^{-1})=(\boldsymbol{\Lambda}_g\boldsymbol{\Lambda}_g^T+\boldsymbol{D}_g)^{-1}$
	
	Hence we conclude that, the variational parameter is essentially the conditional expectation and covariance of $\u_{ig}|\y_i$.

	\section{Parameter estimates for the family of models}\label{family}
	From here, we will derive the family of 8 models by setting different constrains on $\boldsymbol{\Sigma}$. Notice the following identities are easy to verify:
	\[
	\sum_{i=1}^{n}z_{ig}=n_g, \log|(d_g\boldsymbol{I}_K)^{-1}|=\log(d_g^{-K})
	\]
	\[
	\frac{\sum_{i=1}^{n}z_{ig}(\tilde{\m}_{ig}\tilde{\m}_{ig}^T+\tilde{\mV}_{ig})}{n_g}=\boldsymbol{\theta}_g
	\]

	\begin{enumerate}
		\item ``UUU": we do not put any constrain on $\boldsymbol{\Lambda}_g, \boldsymbol{D}_g$. The Solution is exactly the same as above derivation.

		\item ``UUC": we assume $\boldsymbol{D}_g=d_g\boldsymbol{I}_K$, and no constrain for $\boldsymbol{\Lambda}_g$. 
		Except $\boldsymbol{D}_g$, the rest estimation is exactly same as model ``UUU". 
		\[
		\hat{d_g}=\frac{1}{K}tr\{\boldsymbol{\Sigma}_g-2\boldsymbol{\Lambda}_g\boldsymbol{\beta}_g\boldsymbol{S}_g+\boldsymbol{\Lambda}_g\boldsymbol{\theta}_g\boldsymbol{\Lambda}_g^T\}
		\]
		
		\item ``UCU": we assume $\boldsymbol{D}_g=\boldsymbol{D}$, and no constrain for $\boldsymbol{\Lambda}_g$. 
		Except $\boldsymbol{D}_g$, the rest estimation is exactly same as model ``UUU". Take derivative respect to $\boldsymbol{D}^{-1}$, we get:
		\[
		\hat{\boldsymbol{D}}=\frac{1}{n}\sum_{g=1}^{G}n_gdiag\{\boldsymbol{\Sigma}_g-2\boldsymbol{\Lambda}_g\boldsymbol{\beta}_g\boldsymbol{S}_g+\boldsymbol{\Lambda}_g\boldsymbol{\theta}_g\boldsymbol{\Lambda}_g^T\}
		\]
		
		\item "UCC": we assume $\boldsymbol{D}_g=d\boldsymbol{I}_K$, and no constrain for $\boldsymbol{\Lambda}_g$. 
		Except $\boldsymbol{D}_g$, the rest estimation is exactly same as model ``UUU". Follow the same procedure as model ``UUC" and ``UCU", we get:
		\[
		\hat{d}=\frac{1}{Kn}\sum_{g=1}^{G}n_gtr\{\boldsymbol{\Sigma}_g-2\boldsymbol{\Lambda}_g\boldsymbol{\beta}_g\boldsymbol{S}_g+\boldsymbol{\Lambda}_g\boldsymbol{\theta}_g\boldsymbol{\Lambda}_g^T\}
		\]
		
		\item ``CUU": we assume $\boldsymbol{\Lambda}_g=\boldsymbol{\Lambda}$, and no constrain for $\boldsymbol{D}_g$. Except $\boldsymbol{\Lambda}$, the rest estimations are exactly same as model ``UUU". Taking derivative of $l_2$ with respect to $\boldsymbol{\Lambda}$ gives us:
		\[
		\frac{\partial l_2}{\partial\boldsymbol{\Lambda}}=\sum_{g=1}^{G}n_g(\mathbf{D}_g^{-1}\mathbf{S}_g\boldsymbol{\beta}_g^T-\mathbf{D}_g^{-1}\boldsymbol{\Lambda}\boldsymbol{\theta}_g)
		\]
		which must be solved for $\boldsymbol{\Lambda}$ in a row-by-row manner. Let $\lambda_i$ to represent the ith row of $\boldsymbol{\Lambda}$, and $r_i$ to represent ith row of $\sum_{g=1}^{G}n_g(\mathbf{D}_g^{-1}\mathbf{S}_g\boldsymbol{\beta}_g^T)$. Therefore
		\[
		\lambda_i=r_i(\sum_{g=1}^{G}\frac{n_g}{d_{g(i)}}\boldsymbol{\theta}_g)^{-1}
		\]
		where $d_{g(i)}$ is the ith entry of $\mathbf{D}_g$
		
		\item ``CUC": we assume $\boldsymbol{\Lambda}_g=\boldsymbol{\Lambda}, \boldsymbol{D}_g=d_g\boldsymbol{I}_K$. Estimation of $\boldsymbol{\Lambda}_g$ are exactly same as model ``CUU". Estimation of $\boldsymbol{D}_g$ are exactly same as model ``UUC". 
		
		\item ``CCU": we assume $\boldsymbol{\Lambda}_g=\boldsymbol{\Lambda}, \boldsymbol{D}_g=\boldsymbol{D}$. Estimation of $\boldsymbol{\Lambda}_g$ are exactly same as model ``CUU". Estimation of $\boldsymbol{D}_g$ are exactly same as model ``UCU":

		\item ``CCC": we assume $\boldsymbol{\Lambda}_g=\boldsymbol{\Lambda}, \boldsymbol{D}_g=d\boldsymbol{I}_K$. Estimation of $\boldsymbol{\Lambda}_g$ are exactly same as model ``CUU". Estimation of $\boldsymbol{D}_g$ are exactly same as model ``UCC". 
		
	\end{enumerate}

		\section{True Parameters for $\mathbf{\Sigma}_g$ in Simulation Studies}\label{sim:para}
		\footnotesize
		\vspace{-0.1in}
		True $\boldsymbol{\Lambda}$ and $\boldsymbol{D}$ for $\boldsymbol{\Sigma}$ in Simulation Study 1:
\vspace{-0.1in}		
		\[
		\boldsymbol{\Lambda}=\left[\begin{matrix}
-0.003 & 0.386 & -0.242 \\ 
 -0.278 & 0.090 & 0.128 \\ 
-0.131 & 0.187 & 0.375 \\ 
 0.424 & 0.092 & -0.983 \\ 
0.038 & -0.796 & -0.423 \\ 
 0.275 & 0.062 & 0.242 \\ 
 -0.222 & 0.204 & -0.574 \\ 
-0.100 & 0.116 & -0.265 \\ 
 0.284 & 0.422 & -0.205 \\ 
0.030 & -0.353 & 0.153 \\ 
		\end{matrix}\right],\quad \quad  \boldsymbol{D}=0.01*\mathbf{I}_{10}.
		\]
		
		True $\boldsymbol{\Lambda}_g$ and $\boldsymbol{D}_g$ for $\boldsymbol{\Sigma}_g$ in Simulation Study 2:
		\vspace{-0.1in}
				\[
		\boldsymbol{\Lambda}_1=\left[\begin{matrix}
 -0.003 & 0.386 & -0.242 \\ 
 -0.278 & 0.090 & 0.128 \\ 
 -0.131 & 0.187 & 0.375 \\ 
 0.424 & 0.092 & -0.983 \\ 
 0.038 & -0.796 & -0.423 \\ 
 0.275 & 0.062 & 0.242 \\ 
 -0.222 & 0.204 & -0.574 \\ 
 -0.100 & 0.116 & -0.265 \\ 
0.284 & 0.422 & -0.205 \\ 
 0.030 & -0.353 & 0.153 \\ 
		\end{matrix}\right], 
		\boldsymbol{\Lambda}_2=\left[\begin{matrix}
		-0.426 & -0.289 & 0.050 \\ 
		-0.070 & 0.267 & 0.120 \\ 
		0.126 & -0.184 & -0.140 \\ 
		0.276 & -0.690 & 0.394 \\ 
		0.085 & -0.243 & -0.400 \\ 
		-0.137 & 0.104 & -0.305 \\ 
		0.400 & 0.491 & -0.434 \\ 
		0.199 & 0.334 & 0.054 \\ 
		0.167 & 0.022 & -0.167 \\ 
		0.299 & -0.133 & -0.338 \\  
		\end{matrix}\right],
		\boldsymbol{\Lambda}_3=\left[\begin{matrix}
		0.082 & -0.167 & 0.050 \\ 
		0.146 & 0.123 & -0.033 \\ 
		0.164 & -0.075 & -0.142 \\ 
		-0.107 & -0.062 & 0.002 \\ 
		0.086 & 0.054 & -0.143 \\ 
		-0.078 & -0.051 & 0.155 \\ 
		-0.074 & -0.252 & -0.048 \\ 
		-0.059 & 0.112 & 0.076 \\ 
		0.047 & 0.054 & -0.019 \\ 
		0.220 & -0.122 & -0.026 \\  
		\end{matrix}\right]
		\]
				\[\boldsymbol{D}_1=\text{diag}\left[0.03,0.004,0.028,0.015,0.005,0.029,0.003,0.016,0.014,0.015\right]\]
				\vspace{-0.45in}
				\[
		\boldsymbol{D}_2=\text{diag}\left[0.004,0.03,0.015,0.003,0.029,0.015,0.028,0.03,0.005,0.03\right]\]
			\vspace{-0.45in}
		\[
		\boldsymbol{D}_3=\text{diag}\left[0.022,0.006,0.03,0.018,0.011,0.002,0.004,0.015,0.025,0.005\right]
		\]

\end{document}